\documentclass{article}

\usepackage[preprint]{neurips_2026}

\usepackage[utf8]{inputenc} 
\usepackage[T1]{fontenc}    
\usepackage{hyperref}       
\usepackage{url}            
\usepackage{booktabs}       
\usepackage{amsfonts}       
\usepackage{nicefrac}       
\usepackage{microtype}      
\usepackage{xcolor}         
\usepackage{graphicx}
\usepackage{tabularx}
\usepackage{algorithm}
\usepackage{algorithmic}
\usepackage{amsmath}
\usepackage{float}

\title{Ragged Paged Attention: A High-Performance and Flexible LLM Inference Kernel for TPU}

\author{%
  Jevin Jiang \\
  \texttt{jevinjiang@google.com} \\
  \And
  Ying Chen \\
  \texttt{chenyin@google.com} \\
  \AND
  Blake Hechtman \\
  \texttt{blakehechtman@google.com} \\
  \And
  Fenghui Zhang \\
  \texttt{fhzhang@google.com} \\
  \And
  Yarong Mu \\
  \texttt{ymu@google.com} \\
}

\begin{document}

\maketitle
\maketitle
\begin{abstract}
  Large Language Model (LLM) deployment is increasingly shifting to cost-efficient accelerators like Google’s Tensor Processing Units (TPUs), prioritizing both performance and total cost of ownership (TCO). However, existing LLM inference kernels and serving systems remain largely GPU-centric, and there is no well-established approach for efficiently mapping LLM workloads onto TPU architectures—particularly under the dynamic and ragged execution patterns common in modern serving. In this paper, we present Ragged Paged Attention\footnote{Ragged Paged Attention (RPA) is available at \url{https://github.com/vllm-project/tpu-inference/blob/main/tpu_inference/kernels/ragged_paged_attention/v3/kernel.py}} (RPA), a high-performance and flexible attention kernel for TPUs, implemented using Pallas\cite{pallas2023docs} and Mosaic. RPA addresses these challenges through three key techniques: (1) fine-grained tiling to enable efficient dynamic slicing over ragged memory, (2) a custom software pipeline that fuses KV cache updates with attention computation, and (3) a distribution-aware compilation strategy that generates specialized kernels for decode, prefill, and mixed workloads. Evaluated on Llama 3 8B\cite{dubey2024llama} on TPU7x, RPA achieves up to 86\% memory bandwidth utilization (MBU) in decode and 73\% model FLOPs utilization (MFU) in prefill. Integrated as the primary TPU backend in vLLM\cite{kwon2023vllm} and SGLang\cite{zheng2024sglang}, RPA provides a production-grade foundation for efficient TPU inference and offers practical insights into kernel design.
\end{abstract}

\section{Introduction}
The Transformer\cite{vaswani2017attention} architecture, particularly its attention mechanism, forms the foundation of modern Large Language Models (LLMs) and drives the demand for scalable inference systems. As deployment scales, accelerators like Google’s Tensor Processing Units (TPUs) have emerged as attractive options by offering strong performance per total cost of ownership (TCO).

Despite this trend, there is currently no well-established approach for implementing high-performance LLM inference on TPUs in open and production-oriented settings. Existing advances in attention kernel design---most notably the FlashAttention family\cite{dao2022flashattention, dao2023FlashAttention2, shah2024flashattention3, zadouri2026flashattention4}---have been developed almost exclusively for GPUs, with designs increasingly specialized to specific GPU architectures. Conversely, TPU-oriented kernel design remains largely unexplored in the public domain, leaving the efficient mapping of modern LLM workloads onto TPUs an open challenge.

Two challenges underlie this gap. First, XLA’s automated optimization path is often difficult for users to inspect or tune manually when performance fluctuates. Second,  while XLA is highly optimized for the static workloads, whereas today’s LLM serving environments are inherently dynamic. Frameworks like vLLM now frequently mix prefill and decode phases with varying sequence lengths; bridging the gap between these "ragged" patterns and XLA’s static-first architecture is a primary goal of this work. At the same time, TPU memory layouts are tiled and coarse-grained, making fine-grained slicing and KV cache updates non-trivial. Scatter and gather operations can incur additional overhead when access patterns misalign with the underlying layout, and JAX’s immutable array semantics further limit flexible in-place updates. Together,  these factors make it challenging to efficiently support dynamic, ragged workloads.

JAX/XLA ecosystem provides a path forward. Pallas, a JAX kernel language, together with its compiler Mosaic, exposes low-level primitives for direct control of  Direct Memory Access (DMA)/ Remote DMA (RDMA) and register-level computation. This allows developers to rapidly implement custom kernels and workload-specific optimizations, providing a fast time-to-market solution.

\begin{figure}[t] 
  \centering
  \includegraphics[width=0.8\linewidth, trim={0cm 0.5cm 0cm 0cm}, clip]{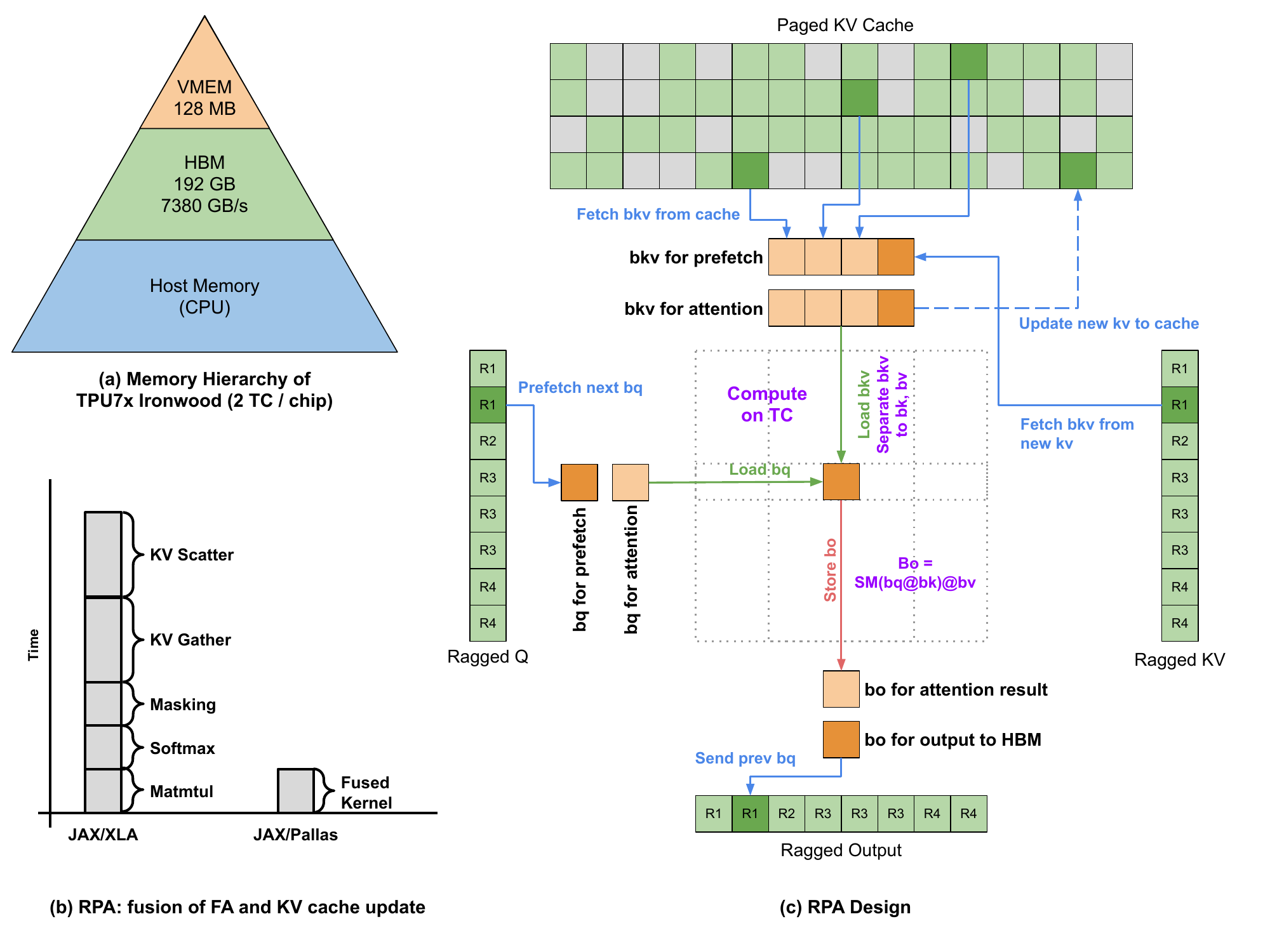}
  \caption{\textbf{(a)} Memory hierarchy of a TPU7x chip, comprising two TensorCores (TCs). Each TC features 64 MB VMEM and 96 GB HBM (chip total: 128 MB VMEM, 192 GB HBM, 7380 GB/s HBM bandwidth). \textbf{(b)} RPA Fusion: RPA integrates KV cache updates into standard FlashAttention fusions, addressing a major TC performance bottleneck. \textbf{(c)}  RPA execution flow: mixed batch (R1-R4) are processed using a double-buffering scheme for Q-blocks, KV-blocks, and output blocks. The KV-blocks are dynamically composed from both the existing paged KV cache and newly projected KV, with asynchronous updates to the KV cache performed during TC computation.}
  \label{fig:rpa_architecture}
\end{figure}

We implement \textbf{Ragged Paged Attention (RPA)} using Pallas to address these challenges via three innovations:
\begin{itemize}
    \item \textbf{Fine-Grained Tiling for Raggedness:} enforces XLA to choose the smallest tiling and enables efficient dynamic slicing over ragged memory buffers.
    \item \textbf{KV-update-fused Pipeline:} effectively hides KV cache update latency by overlapping it with attention computation, eliminating TensorCore scatter overheads commonly incurred in today’s JAX/XLA path.
    \item \textbf{Distribution-Aware Compilation:} dispatches specialized kernels based on sequence length regimes, optimizing performance for different workloads.
\end{itemize}

RPA delivers high performance across diverse LLMs; for example, on Llama 3 8B, it achieves up to $86\%$ memory bandwidth utilization (MBU) in decode (memory-bound) and $73\%$ model FLOPs utilization (MFU) in prefill (compute-bound) on TPU7x. 

Beyond performance, RPA is highly flexible across: (1) \textbf{Models:} supports a wide range of models out of the box, including different sharding strategies, quantization schemes, and attention variants (MHA\cite{vaswani2017attention}, MQA\cite{shazeer2019fast}, GQA\cite{ainslie2023gqa}); (2) \textbf{TPU Generations:} runs on TPU v4–v7 requiring only block-size tuning; (3) \textbf{Workloads:} efficiently handles decode, prefill, and mixed batches.

Together, RPA bridges the gap between modern LLM serving and TPU hardware, delivering practical performance gains and serving as a production-grade, TPU-native kernel foundation. It is integrated as the primary TPU backend in leading LLM serving frameworks, including vLLM and SGLang, providing both efficient large-scale LLM inference and practical insights into TPU kernel design.
\section{Background}

\subsection{TPU Architecture}
\subsubsection{TensorCore and SparseCore}

The TPU architecture exposes two types of programmable cores: TensorCore (TC) and SparseCore (SC), each optimized for different classes of workloads.

\begin{itemize}
    \item \textbf{TensorCore} serves as the primary compute engine, optimized for dense matrix operations, including matrix multiplications and element-wise operations on large tensors.
    \item \textbf{SparseCore} is designed for irregular and sparse workloads, supporting fine-grained operations such as gather, scatter, and sorting. Additionally, it provides a clean heterogeneous hardware architecture for offloading.

\end{itemize}

In this work, we focus on TC, as it is currently the primary compute engine for TPU LLM inference.

\begin{table}[ht]
\centering
\caption{Key Components of TensorCore in TPU7x}
\label{tab:tpu_arch}
\begin{tabularx}{\textwidth}{@{} l X @{}}
\toprule
\textbf{Component} & \textbf{Description} \\ \midrule
HBM  & Off-chip high-bandwidth memory. \\
VMEM & On-chip vector memory with direct load/store to VREG. \\
SMEM & On-chip scalar memory with direct load/store to SREG. \\
VREG & Vector register of shape ($8 \times 128$) with 32-bit entries. \\
SREG & Scalar register with 32-bit entries. \\
VPU  & Vector Programmable Unit. A 128 lane $\times$ 8 sublane SIMD processor in the TensorCore. \\
MXU  & Matrix Unit. A $256\times256$ Systolic Array used primarily for dense matrix-matrix multiplication. \\ \bottomrule
\end{tabularx}
\end{table}

\begin{figure}[ht] 
  \centering
  \includegraphics[width=0.9\linewidth, trim={0cm 14cm 0cm 0cm}, clip]{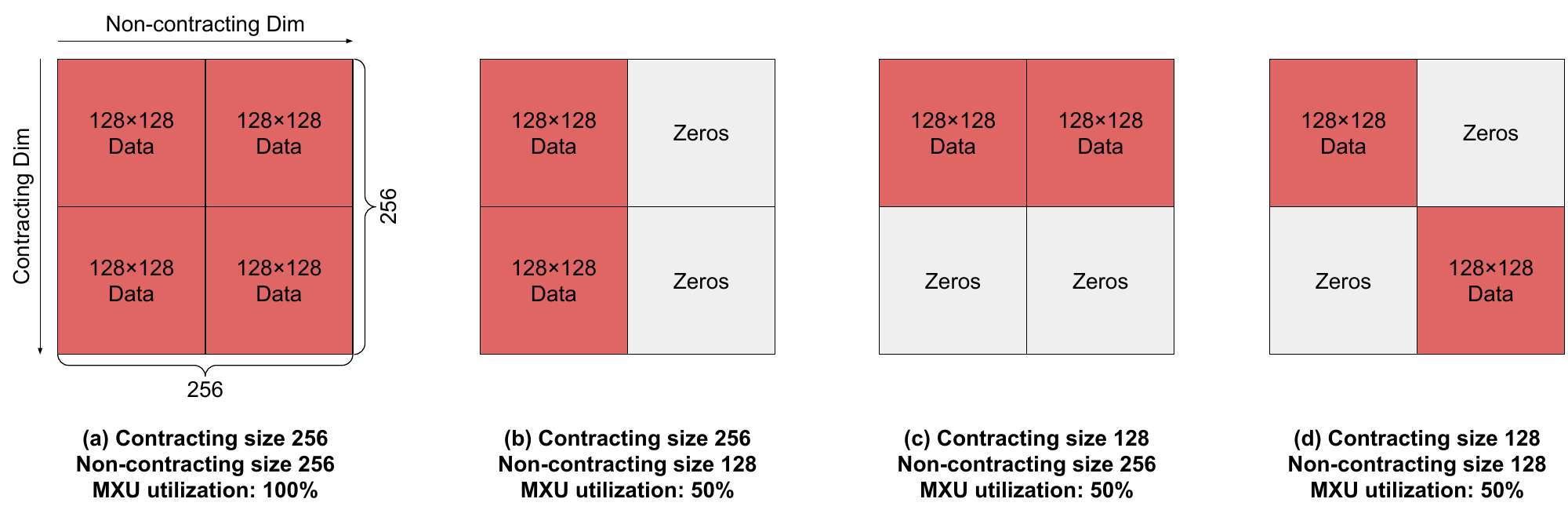}
  \caption{\textbf{(a)(b)(c)(d)} illustrate the Matrix Unit (MXU) data mapping strategy to achieve maximum utilization under different cases on TPU7x }
  \label{fig:mxu}
\end{figure}

The MXU is the primary source of compute throughput on TPUs, delivering high FLOPs for dense matrix multiplication. On TPU7x, it operates on a $256 \times 256$ systolic array, with the right-hand side (RHS) mapped and reused across the computation. As a result, utilization is sensitive to operand shapes: when either the contracting or non-contracting dimension is smaller than the MXU size, the array is underutilized. For example, in the Llama 3 8B model, the attention head dimension is $128$, which limits $2$ required matmuls in attention to at most $50\%$ MXU utilization.

\subsubsection{Memory Layout and Packing}
\label{sec:memory_layout}

Data residing within the TPU memory hierarchy—including HBM, VMEM, and SMEM—is organized into a tiled layout. This tiling ensures alignment with the hardware's DMA granularity, which typically operates on 512-byte words, equivalent to 128 contiguous 32-bit entries.

\begin{figure}[ht] 
  \centering
  \includegraphics[width=1.0\linewidth, trim={0cm 11.6cm 0cm 0cm}, clip]{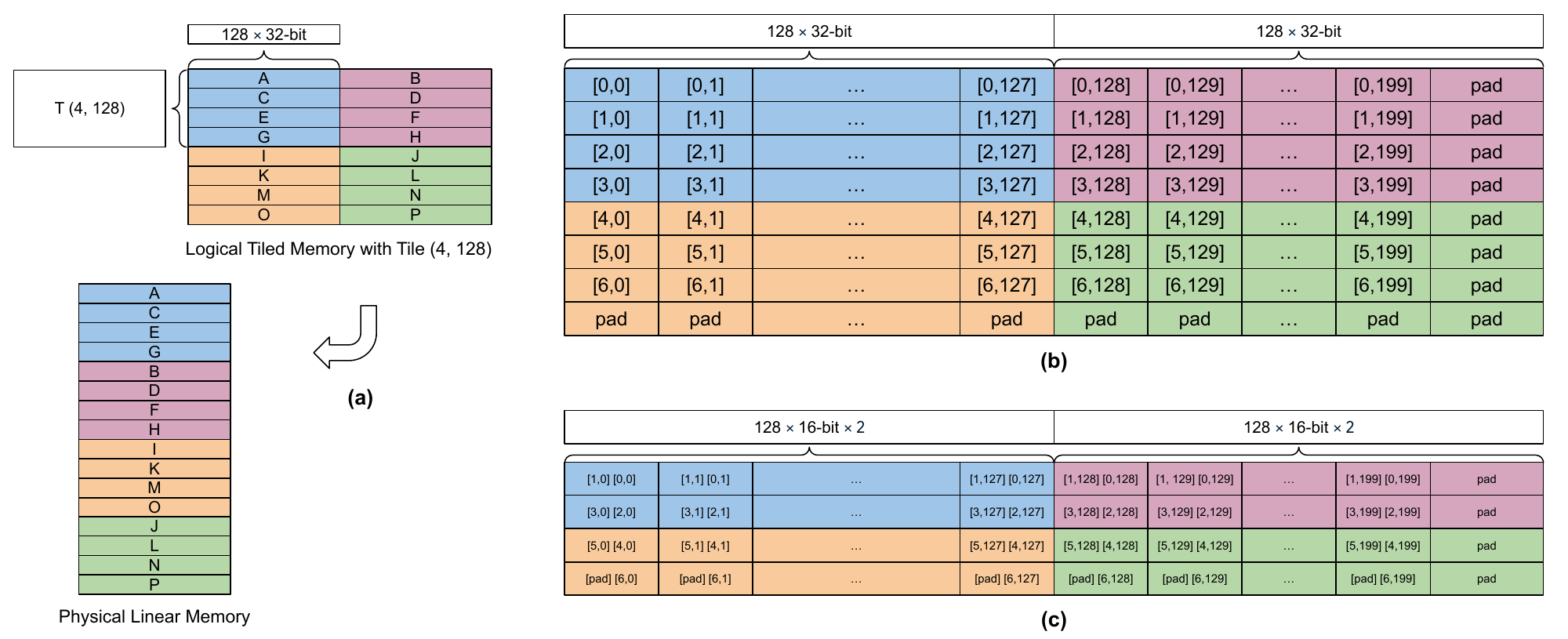}
  \caption{\textbf{(a)} \texttt{F32(8,256)} with \texttt{T(4,128)} in logical tiled memory vs. physical Linear memory. \textbf{(b)} \texttt{F32(7, 200)} with \texttt{T(4, 128)} in logical tiled memory \textbf{(c)} \texttt{BF16(7, 200)} with \texttt{T(4, 128)} in logical tiled memory}
  \label{fig:tpu_memory}
\end{figure}

As illustrated in Figure~\ref{fig:tpu_memory}(a), consider a tensor with a logical shape of \texttt{F32(8,256)} and a tile shape \texttt{T(4,128)}. Each tile is composed of four indivisible "chunks" with each chunk containing 128 elements to match the typical DMA width in TC. While the logical representation follows a standard row-major order: ABCDEFGHIJKLMNOP, the physical linear memory allocation follows a tiled sequence: ACEGBDFHIKMOJLNP.

For narrow data types (bit-width $< 32$), TPUs employ \textit{packing} to improve memory efficiency. Multiple elements from adjacent rows of the logical tensor are packed into a single 32-bit ``element,'' where lower and higher bits correspond to different rows.

To illustrate the impact of tiling and packing in memory, consider a tensor with a logical shape of \texttt{F32(7,200)} utilizing a tile size of \texttt{T(4,128)}:

\begin{itemize}
    \item \texttt{F32} Representation in Figure ~\ref{fig:tpu_memory}(b): In a standard 32-bit format, the tensor requires implicit padding to align with the \texttt{T(4, 128)} tiling boundary. Consequently, the physical memory footprint expands to \texttt{F32(8, 256)} to accommodate the minimum tiling requirements.
    \item \texttt{BF16} Representation in Figure ~\ref{fig:tpu_memory}(c): With a bit-width of 16, the packing factor is 2. The TPU packs the high 16 bits from row $2i+1$ and the low 16 bits from row  $2i$  into a single 32-bit packed element. While the same implicit padding is applied to align with the tile boundaries, the physical memory footprint expands to \texttt{BF16(8, 256)}. Compared to  \texttt{F32}, the packed nature of the data results in a 50\% reduction in total memory consumption.
\end{itemize}

The discrepancy between logical and physical layouts, together with packing for narrow data types, presents a fundamental challenge in arbitrary memory slices on tiled dimensions, particularly writing ragged input directly to memory without blending on VREGs. In the context of RPA, the kernel is explicitly designed to account for these constraints in order to enable efficient dynamic slicing and scatter/gather operations. By placing ragged dimensions along non-tiled axes and inserting packing as the second minor dimension, RPA enforces XLA to use the minimal tile \texttt{T(packing, 128)} and enables arbitrary and dynamic slice.

\subsection{Compiler and Custom Kernel: Pallas/Mosaic}

\begin{figure}[ht] 
  \centering
  \includegraphics[width=1.0\linewidth, trim={0cm 8cm 0cm 0cm}, clip]{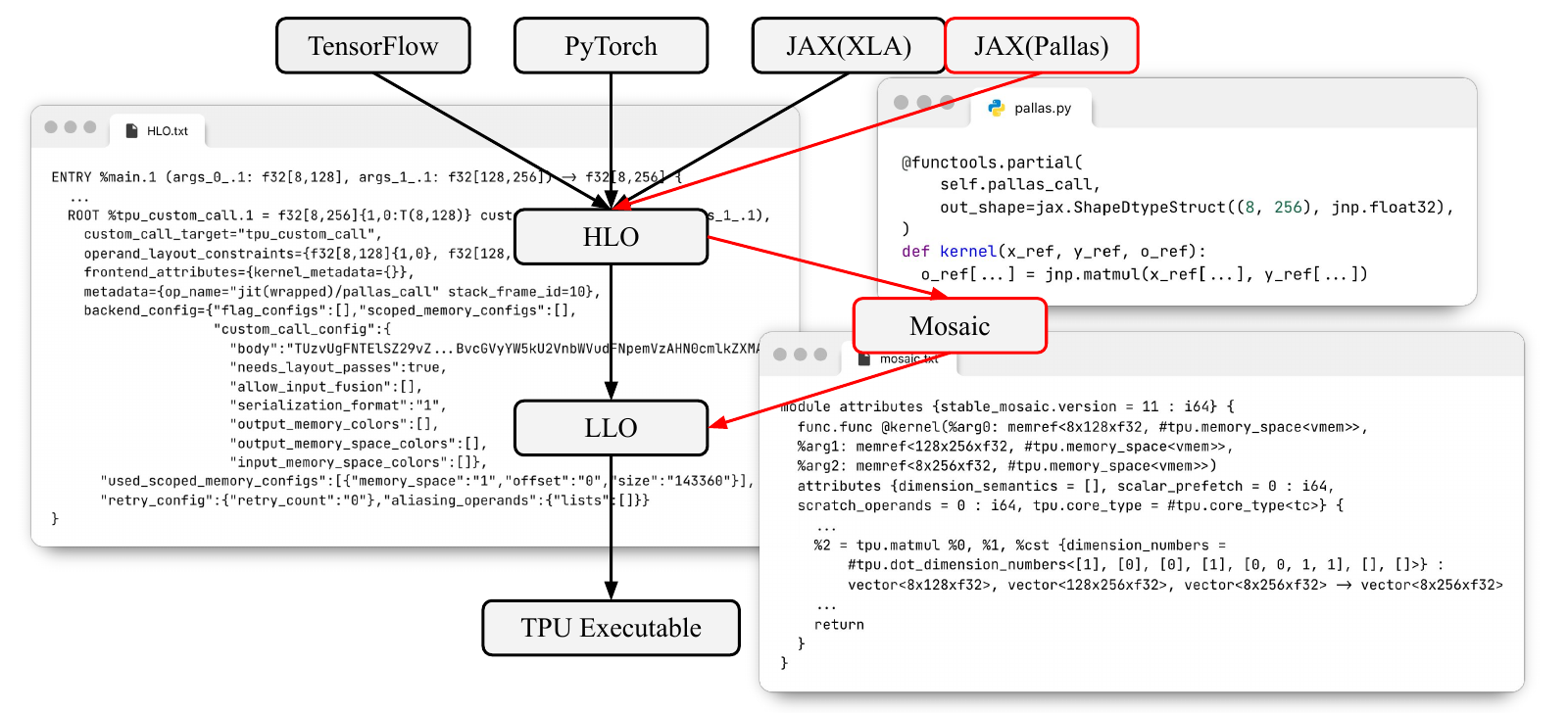}
  \caption{Pallas Lowering}
  \label{fig:lowering}
\end{figure}

Pallas is a kernel programming extension to JAX that enables users to write custom TPU kernels while reusing familiar JAX APIs at a lower level of abstraction. It provides explicit control over memory and compute, including primitives for asynchronous DMAs/RDMAs across different memory spaces or devices and load/store/compute with VREGs.

Mosaic is the MLIR-based compiler backend for Pallas, providing an “escape hatch” from the standard XLA compilation pipeline. Instead of being fully lowered through XLA, a Pallas kernel is encapsulated as a custom HLO operation with encoded metadata. During lowering, Mosaic extracts this metadata, translates the kernel into MLIR, and performs further lowering and optimization. Ultimately, both the standard XLA path and the Mosaic path converge to the same low-level operations (LLO), which is close to machine code, where final optimizations are applied before generating the executable.

\subsection{Software Pipeline}

\begin{figure}[ht] 
  \centering
  \includegraphics[width=0.8\linewidth, trim={0cm 9cm 0cm 0cm}, clip]{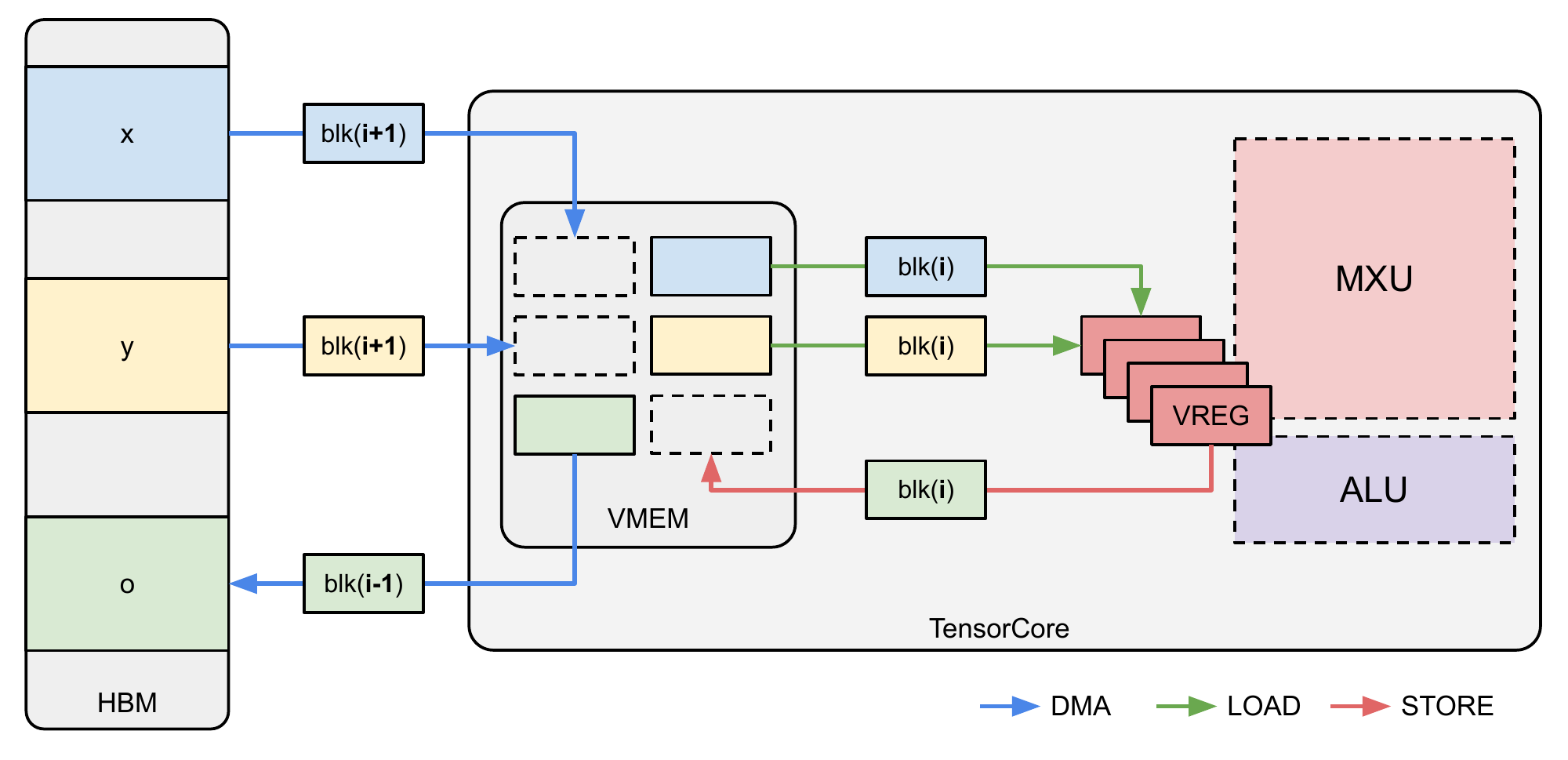}
  \caption{Double-buffered TPU pipeline: compute on \texttt{x\_block($i$)}, \texttt{y\_block($i$)}, \texttt{o\_block($i$)} overlaps with prefetch of \texttt{x\_block($i+1$)}, \texttt{y\_block($i+1$)} and write-back of \texttt{o\_block($i-1$)}.}
  \label{fig:pipeline}
\end{figure}

TPUs utilize a single-threaded execution model but achieve parallelism through asynchronous DMA operations, which enable the overlapping of TensorCore computation with DMAs. This overlap is the primary mechanism for extracting hardware parallelism on TPUs. While Pallas provides high-level abstractions---such as block specifications---to automatically generate multi-buffered pipelines, these often abstract away low-level DMA primitives and semaphore management (e.g., \texttt{async\_copy} and \texttt{semaphore\_signal}).

In RPA, we depart from these high-level abstractions and explicitly implement a custom pipelining strategy using Pallas's primitive APIs. By retaining granular control over data movement and compute scheduling, we can more tightly integrate KV cache updates directly into the attention kernel. This design enables more aggressive fusion and eliminates the limitation associated with generic compiler-generated pipelines.

\subsection{vLLM}
\subsubsection{PagedAttention}
PagedAttention is a key innovation in vLLM for managing the KV cache using a paged abstraction, analogous to virtual memory systems. Instead of requiring contiguous allocation, it partitions each sequence’s KV cache into fixed-size pages, allowing keys and values to be stored in non-contiguous memory. This allows vLLM to significantly increase its memory efficiency and avoid fragmentation to accommodate larger batch sizes and/or longer context windows. During attention computation, the kernel dynamically gathers the required KV blocks based on page mappings.

While effective for memory efficiency and flexibility, this design introduces challenges for TPU execution. First, KV accesses require gathering data from dynamically computed, non-contiguous memory addresses, which complicates efficient DMA scheduling. Second, KV cache updates require scattering newly generated key-value pairs into partially filled blocks, particularly during autoregressive decoding where updates occur at single-token granularity. These irregular memory access patterns are not well aligned with the TPU compiler stack, which is optimized for statically shaped workloads.

\begin{figure}[ht] 
  \centering
  \includegraphics[width=1.0\linewidth, trim={0cm 15cm 0cm 0cm}, clip]{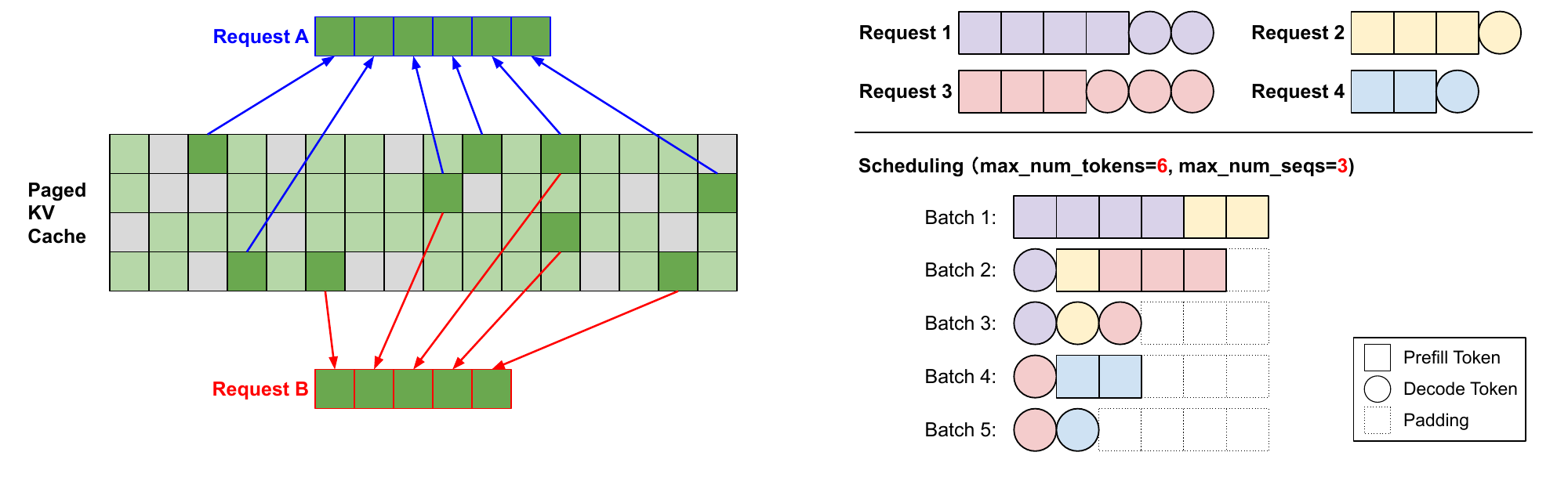}
  \caption{\textbf{Left}: Illustration of a paged KV cache, where the key-value (KV) tensors for different requests (e.g., request A and request B) are stored across non-contiguous memory pages. \textbf{Right}: Mixed-batch scheduling produces highly ragged inputs with varying sequence lengths, often requiring padding for alignment.}
  \label{fig:vllm}
\end{figure}

\subsubsection{Mixed Batch}
The vLLM scheduler eliminates the traditional separation between prefill and decode phases by treating prompt tokens and generated tokens uniformly within a single execution model.

While this unified abstraction simplifies the serving pipeline, it significantly complicates TPU kernel design. In particular, it produces inputs that are both ragged, due to variable sequence lengths, and dynamic, as the workload distribution is only known at runtime. Such characteristics are difficult to handle efficiently on TPUs, where compilation and execution are optimized for more regular and statically shaped workloads.

\section{Ragged Paged Attention}

\begin{table}[ht]
\caption{RPA Notations}
\label{tab:rpa_notations}
\centering
\small
\begin{tabularx}{\textwidth}{@{} l X @{}}
\toprule
\textbf{Symbol} & \textbf{Description} \\ \midrule
$Q$ & Query matrix for attention\\
$K$ & Key matrix for attention\\
$V$ & Value matrix for attention\\
$KV$ & Merged $K$ and $V$\\
$s$ & Maximum number of tokens in a mixed batch (flattened across all sequences); includes padding.\\
$n$ & Maximum number of sequences can be scheduled in one mixed batch. \\
$h_q$ & Number of query heads. \\
$h_{kv}$ & Number of key/value heads. \\
$h_g$ & Number of query heads per key/value head (group size). $h_g = h_q / h_{kv}$. \\
$d_k$ & Head dimension for $Q$, $K$, and $V$ (assuming $d_k=d_v$). \\
$p_q$ & Packing factor for $Q$. \\
$p_{kv}$ & Packing factor for $KV$. \\
$b_q$ & Block size along the query sequence length for a single sequence. \\
$b_{kv}$ & Block size along the key/value sequence length for a single sequence. \\
$c_q$ & Compute block size corresponding to $b_q$. \\
$c_{kv}$ & Compute block size corresponding to $b_{kv}$. \\
$B_q$ & The $Q$ block buffer in VMEM. \\
$B_o$ & The attention output block buffer in VMEM. \\
$B_{kv}$ & The $KV$ block buffer in VMEM. \\
$U_{kv}$ & The slice of $B_{kv}$ containing newly projected $KV$ data designated for the cache update.\\
$C_q$ & The $Q$ compute block loaded from $B_q$ to VREGs. \\
$C_o$ & The output from attention with compute blocks. \\
$C_{kv}$ & The $KV$ compute block loaded from $B_{kv}$  to VREGs.\\
$C_k$ & The $K$ compute block separated from $C_{kv}$ and reconstructed in VREGs.\\
$Cv$ & The $V$ compute block separated from $C_{kv}$ and reconstructed in VREGs.\\  \bottomrule
\end{tabularx}
\end{table}

\subsection{Preprocessing}
\subsubsection{The Packing Dimension}
One limitation of current XLA and Pallas/Mosaic is the inability for users to explicitly control memory tiling. For example, given a tensor \texttt{F32(32, 128)}, XLA cannot be forced to use a tile of \texttt{T(1, 128)} and instead selects its own tile, e.g., \texttt{T(8, 128)}. The memory layout of inputs and outputs is fully determined by XLA’s \textit{Layout Assignment} pass in HLO, even for custom kernels. The layout is chosen based on tensor shape and data type, which dictates both tiling and padding. This raises the question: why does tiling matter for RPA?

In standard attention, after projection, $Q$ has shape $(s, h_q, d_k)$, while $K$ and $V$ have shape $(s, h_{kv}, d_k)$. Since $h_{kv}$ is a loop dimension that does not participate fully in matrix multiplication, it is natural to transpose $K$ and $V$ to $(h_{kv}, s, d_k)$ on TPU. According to Section ~\ref{sec:memory_layout}, the last two dimensions of tensors in TPU are tiled by XLA and we need to slice within tile if $h_{kv}$ present as the second minor.

For mixed batches, the flattened token dimension s concatenate dynamic lengths from multiple sequences. Placing it in the second minor dimension is inefficient: it requires computing the effective range inside each tile, and in many cases, multiple rows are packed together. This necessitates unpacking, blending, and repacking during dynamic slicing, which is expensive on the VPU. To avoid this overhead, it is critical that ragged and dynamic dimensions are not assigned to the tiling dimensions (the last two dimensions). This allows the kernel to dynamically slice the effective size based on runtime needs.

Another challenge arises when sharding along the head dimensions $h_q$ and $h_{kv}$. Although RPA supports a wide range of attention types---including MHA, MQA, and GQA, which vary in how $Q$ heads share a $KV$ head---$h_q$ and $h_{kv}$ can be arbitrary. \textit{Tensor Parallelism} (TP) sharding of $h_{kv}$ and the use of quantized data types can further reduce the per-device $h_{kv}$ to values smaller than the minimum tile size, \texttt{T(packing, 128)}, causing XLA to introduce implicit padding. For example, a \texttt{BF16(12, 128)} tensor is padded to \texttt{BF16(16, 128)} by XLA to align with a \texttt{T(8, 128)} tile; however, utilizing a smaller configuration such as \texttt{T(2, 128)} can avoid this padding overhead.

To support effective slicing and mitigate padding, we introduce a \textit{packing dimension} in the second minor dimension of $Q, K,$ and $V$. This enforces XLA to select the minimum tile \texttt{T(packing, 128)}.

\subsubsection{Q Reshape}
After introducing the packing dimension, the shape of $Q$ becomes $(s, \lceil h_q / p_q \rceil, p_q, d_k)$, which can also be expressed as $(s, \lceil h_{kv} \cdot h_g / p_q \rceil, p_q, d_k)$. Since $h_{kv}$ serves as the loop dimension, we further transpose $Q$ to place it in the leading position, yielding $(h_{kv}, s, \lceil h_g / p_q \rceil, p_q, d_k)$. This layout enables more efficient utilization of the MXU during attention computation. Specifically, when loading a $Q$ compute block $C_q$ into VREGs, we reshape it into $(c_q \cdot h_g, d_k)$, allowing multiple query heads within a group to be processed against the same $C_k$ and $C_v$. As a result, the shared $C_k$ and $C_v$ operands can remain resident in the MXU across all grouped $C_q$, reducing the overhead of repeatedly loading them into the MXU. Even though it is done differently in TPU, this is a common GQA optimization to improve effective matrix multiplication throughput (FLOPs/s).

However, for $K$ and $V$, it is not necessary to explicitly move $h_{kv}$ to the leading dimension, despite it being a loop dimension. This is because the required transposition can be fused into the data loading path, as discussed in Section ~\ref{sec:kv_xpos_fuse}.

\subsubsection{KV Merge}
Both $K$ and $V$ share the same logical shape $(s_{kv}, \lceil h_{kv} / p_{kv} \rceil, p_{kv}, d_k)$. For \texttt{BF16} ($p_{kv}=2$), in a common case where $h_{kv}=1$ per device (e.g., after tensor parallel sharding), implicit padding is still introduced due to tiling constraints. To mitigate this inefficiency, we merge $K$ and $V$ along the head dimension, resulting in a merged representation with shape $(s_{kv}, \lceil 2h_{kv} / p_{kv} \rceil, p_{kv}, d_k)$.

This merged $KV$ representation provides several benefits beyond reducing padding overhead. First, it can reduce the number of vector load and store operations by approximately half, as $K$ and $V$ are fetched and written in a single combined access. This is beneficial because load/store operations are more expensive than typical VPU arithmetic operations. Second, it increases the effective DMA transfer size. For example, in the decode regime, instead of issuing two small DMA transfers (one for $K$ and one for $V$ per token), the merged layout enables a single larger DMA. This not only amortizes the DMA base latency but also makes it easier to satisfy the minimum efficient DMA granularity, thereby improving bandwidth utilization and overall DMA efficiency.

\begin{figure}[ht] 
  \centering
  \includegraphics[width=0.8\linewidth, trim={0cm 12cm 0cm 0cm}, clip]{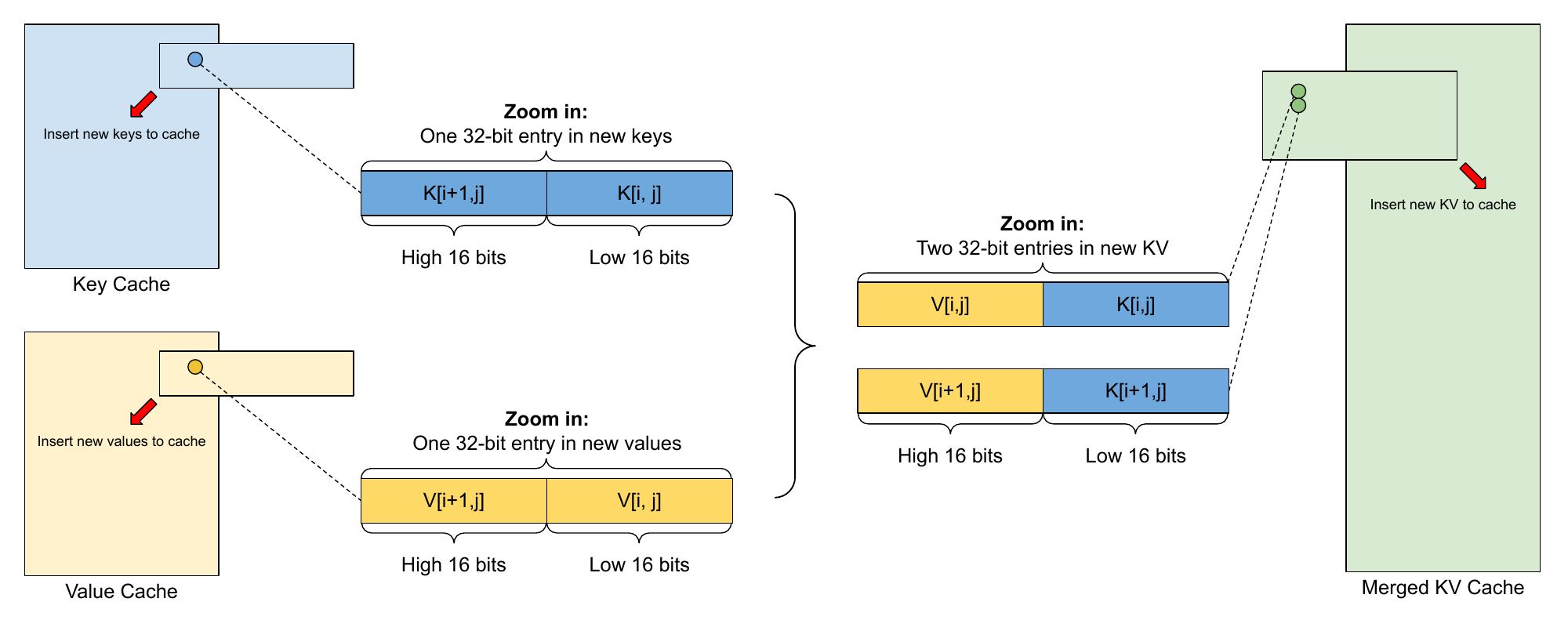}
  \caption{Interleaved packing of K and V to a merged KV representation, such that any slice of the KV cache contains both K and V.}
  \label{fig:kv_merge}
\end{figure}

Separating the merged $C_{kv}$ into $C_k$ and $C_v$ requires loading a $32$-bit packed value, followed by unpacking and repacking into the target data type. When this process does not interfere with MXU scheduling, its overhead can be effectively hidden. However, for lower-precision formats such as FP8, a single $32$-bit word contains two merged $KV$ elements. In this case, iterating over $h_{kv}$ requires repeatedly extracting $8$-bit values from the same $32$-bit $c_{kv}$ to reconstruct $c_k$ and $c_v$ for 2 $h_{kv}$ separately, introducing additional overhead. This trade-off is acceptable in our design, as the $KV$ cache update granularity accounts for $h_{kv}$ to increase transfer efficiency. In other words, the smallest update unit (a single-token slice) contains both $K$ and $V$ for all $KV$ heads.

\subsubsection{Offline Preprocessing and Projection}
\label{sec:projection}

In principle, the projection weight $W_Q$ has shape $(d_{model}, h_q, d_k)$ and both $W_K$ and $W_V$ have shape $(d_{model}, h_{kv}, d_k)$. Those weights could be reshaped and merged offline to match our desired kernel input layouts. Specifically, $W_Q$ could be transformed to $(h_{kv}, \lceil h_g / p_q \rceil, p_q, d_{model}, d_k)$, while $W_K$ and $W_V$ could be interleaved into a merged format of $(\lceil 2h_{kv} / p_{kv} \rceil, p_{kv}, d_{model}, d_k)$. Performing these transformations offline would allow a single \texttt{einsum} during projection to directly produce the query ($Q$) and merged $KV$ tensors in RPA desired shapes.

However, our empirical evaluations revealed that XLA’s \textit{Layout Assignment} pass often fails to preserve the required \texttt{T(packing, 128)} tiling for our kernel. While XLA typically respects a descending minor-to-major layout for the boundary inputs and outputs of a Mosaic custom kernel, it frequently reorders the layout of intermediate tensors---effectively performing implicit transpositions---to optimize for general compute efficiency elsewhere in the HLO graph. This behavior overrides user-specified layout constraints in current XLA versions.

Consequently, to guarantee layout integrity, we defer these transformations to the RPA preprocessing stage rather than offloading them to offline weight preparation. Looking forward, three alternative optimization strategies remain viable:

\begin{itemize}
    \item \textbf{Projection Kernel:} Implementing the projection within a custom kernel that strictly constrains memory layouts. However, this approach cannot guarantee that subsequent operations between the projection and the RPA kernel will not trigger a layout reordering within XLA.
    \item \textbf{Kernel Fusion:} Fusing the linear projection directly into the RPA kernel to ensure that the resulting layout is never altered by the compiler.
    \item \textbf{Layout Propagation:} Should future XLA iterations provide stronger guarantees for propagating user-defined layouts from weights through to custom kernels, the current preprocessing overhead could be entirely eliminated.
\end{itemize}

\subsection{KV Transpose Fusion}
\label{sec:kv_xpos_fuse}

Since $h_{kv}$ is a loop dimension that does not participate in the matrix multiplication, we fuse the required transpose of $KV$ into the data loading path using strided vector loads. After fetching a $KV$ block with shape $(b_{kv}, \lceil 2h_{kv} / p_{kv} \rceil, p_{kv}, d_k)$ from HBM (with global shape $(s_{kv}, \lceil 2h_{kv} / p_{kv} \rceil, p_{kv}, d_k)$) into VMEM, we first reinterpret the buffer and apply a \texttt{shapecast} to obtain a logical view of shape $(b_{kv} \cdot 2h_{kv}, d_k)$. During subsequent compute iterations, we then perform strided vector loads of shape $(c_{kv}, d_k)$ with stride $2h_{kv}$, effectively materializing the transposed layout on-the-fly without introducing an explicit transpose operation.

However, arbitrary strided access patterns can easily introduce bank conflicts in VMEM, leading to inefficient vector load scheduling and degraded performance. To mitigate this, we introduce a small offset in the $KV$ block layout in VMEM such that the effective stride used during vector loads becomes an odd number. This optimization allows the DMA to maintain an effective transfer size while ensuring that the $KV$ loads are efficiently scheduled.

\subsection{KV Cache Update Fusion}

While the TPU follows a single-threaded execution model, it functions similarly to other modern accelerators by leveraging asynchronous DMA operations to overlap data movement with TensorCore computation. To maximize performance, RPA utilizes Pallas's primitive APIs to implement a custom software pipeline. This granular control is necessitated by three primary requirements of the RPA kernel:

\begin{itemize}
    \item \textit{Dynamic Memory Addressing:} The need to fetch $KV$ blocks from non-contiguous, paged memory addresses.
    \item \textit{Ragged Dimension Handling:} The requirement to dynamically calculate and fetch the effective sequence lengths for both $Q$ and $KV$.
    \item \textit{Kernel Fusion:} The integration of $KV$ cache updates (scatter operations) directly into the FlashAttention compute loop to eliminate independent update latency.
\end{itemize}

The pipeline manages four types of DMA operations:
\begin{itemize}
    \item \textbf{Fetch $B_q$}: Retrieve a block of $Q$ from HBM to VMEM by transferring only the effective token range, leaving any remaining regions of the block untouched.
    \item \textbf{Fetch $B_{kv}$}: Gather $KV$ pages from HBM according to the page table, incorporating newly projected $KV$ entries when applicable, and concatenate them into a contiguous $KV$ block in VMEM.
    \item \textbf{Update $U_{kv}$}: Write newly projected $KV$ entries back to the $KV$ cache in HBM.
    \item \textbf{Send $B_o$}: Transfer the output block, corresponding to the effective token range, from VMEM scratch space back to HBM.
\end{itemize}

These DMA operations are overlapped with the following compute stages:
\begin{enumerate}
    \item Load a compute block $C_q$ from $B_q$ into VREGs.
    \item Perform strided loads of $C_{kv}$ from $B_{kv}$, fusing the transpose as described in Section 3.2.
    \item Separating $C_{kv}$ by unpacking and repacking to obtain $C_k$ and $C_v$.
    \item Run \texttt{FlashAttention-2}($C_q, C_k, C_v$). Our implementation adapts Algorithm 1 from FlashAttention-2~\cite{dao2023FlashAttention2}.
    \item Store the resulting output $C_o$ into the output block VMEM buffer $B_o$.
\end{enumerate}

\begin{figure}[ht] 
  \centering
  \includegraphics[width=0.9\linewidth, trim={0cm 15.5cm 0cm 0cm}, clip]{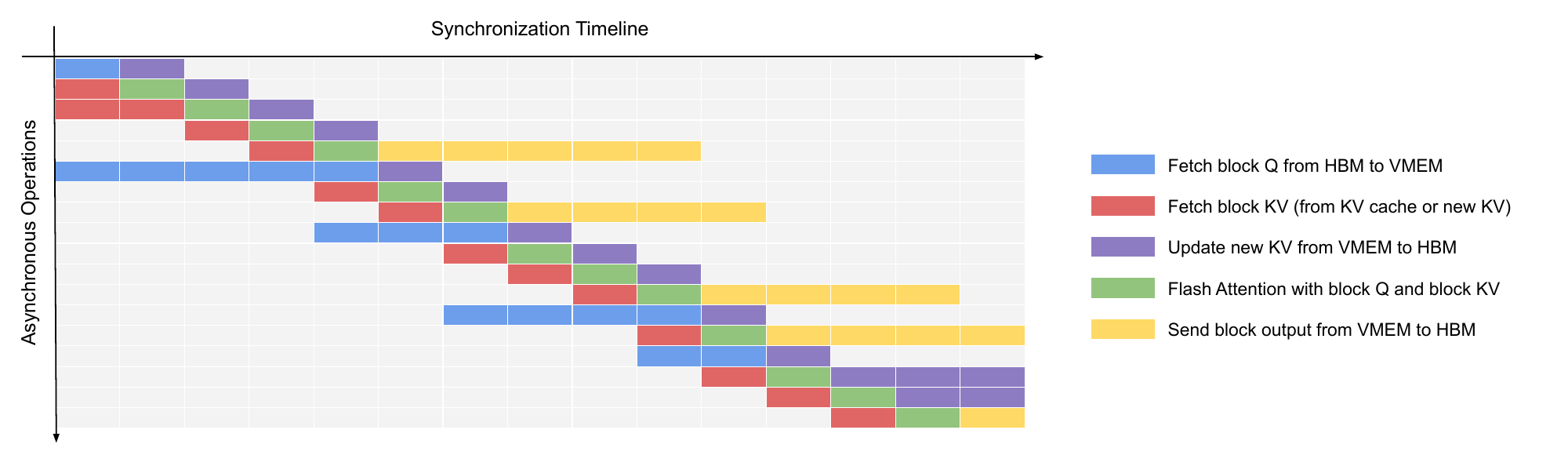}
  \caption{RPA Pipeline Visualization. For each DMA block, the left edge indicates when the asynchronous DMA operation is started, and the right edge indicates when the operation is awaited.}
  \label{fig:rpa_pipeline}
\end{figure}

\subsection{Distribution-aware Compilation}

Although RPA issues effective-size DMA operations---eliminating wasted bandwidth---computation on the TensorCore still requires a static number of VREGs to unroll a compute block, which can lead to unavoidable wastes in attention computation. To mitigate this, we introduce a \textit{distribution-aware compilation} strategy. Specifically, we pass dynamic indices $[i, j, k]$ to the kernel to indicate workload segmentation: sequences $[0, i-1]$ correspond to decode-only requests (token length = 1), sequences $[i, j-1]$ correspond to fixed-size-chunk prefill (where prefill length is static), and sequences $[j, k-1]$ correspond to mixed-batch requests. 

With this distribution information, for each range of sequences, RPA receives a \texttt{case} enum indicating the workload type (decode-only, prefill-only, mixed) and dispatches to the corresponding compiled kernels. In the first two cases, the query block size $b_q$ is known at compile time, allowing the kernel to statically determine VREG usage and schedule computation efficiently. By contrast, in the mixed case, $b_q$ must accommodate unknown effective lengths, potentially leading to wasted compute resources due to padding. To improve scheduling efficiency, we can apply a post-scheduling reordering algorithm to ensure that all decode-only and prefill-only requests are placed at the front of the batch.

Effectively loading $B_{kv}$ for the mixed workload remains challenging, but it is feasible when the model employs a static sliding window for attention. The current RPA implementation does not yet support this optimization.

\begin{algorithm}
\caption{Ragged Paged Attention (RPA) Forward Pass with KV Cache Update Fusion}
\label{alg:rpa_forward}
\begin{algorithmic}[1]
\STATE Start DMA to fetch $B_q^{(r_{start}, 0)}$ and $B_{kv}^{(r_{start}, 0)}$ from HBM to local VMEM scratch
\FOR{$r_{start} \le r < r_{end}$}
    \STATE Load dynamic $Q$ length $l_q$ and $KV$ length $l_{kv}$ for sequence $r$
    \STATE Divide $Q$ into $T_q = \lceil l_q / b_q \rceil$ blocks and $KV$ into $T_{kv} = \lceil l_{kv} / b_{kv} \rceil$ blocks
    \FOR{$0 \le i < T_q$}
        \STATE On chip, initialize: $B_o^{(r, i)} \gets \mathbf{0}$, $l \gets \mathbf{0}$, $m \gets -\infty$
        \STATE On chip, calculate next indices $(r', i')$: if $i+1 < T_q$ then $(r, i+1)$ else $(r+1, 0)$
        \STATE Start DMA for fetching $B_q^{(r', i')}$ when $r' < r_{end}$
        \STATE Wait for DMA fetching $B_q^{(r, i)}$
        \FOR{$0 \le j < T_{kv}$}
            \STATE On chip, calculate next indices $(r'', j'')$: if $j+1 < T_{kv}$ then $(r, j+1)$ else $(r+1, 0)$
            \STATE Start DMA for fetching $B_{kv}^{(r'', j'')}$ when $r'' < r_{end}$ (contains new and cached $KV$)
            \STATE Wait for DMA fetching $B_{kv}^{(r, j)}$
            \STATE Start DMA for updating new $KV$ entries $U_{kv}^{(r, j)}$ to KV cache in HBM when $i+1 = T_q$
            \STATE On chip, calculate dynamic effective block size $b'_{kv} \le b_{kv}$
            \STATE Divide $B_{kv}$ into $N_{kv} = \lceil b'_{kv} / c_{kv} \rceil$ and $B_q$ into $N_q = \lceil b_q / c_q \rceil$ compute blocks
            \FOR{$0 \le x < N_{kv}$}
                \FOR{$0 \le y < N_q$ \textbf{unrolled}}
                    \FOR{$0 \le h < h_{kv}$ \textbf{unrolled}}
                        \STATE Strided load $C_{kv}^{(x, h)}$ to VREGs; separate to $C_k^{(x, h)}$ and $C_v^{(x, h)}$
                        \STATE Load $C_q^{(y, h)}$, $C_o^{(y, h)}$, $l^{(y, h)}$, $m^{(y, h)}$ into VREGs
                        \STATE $S \gets C_q^{(y, h)} (C_k^{(x, h)})^T \in \mathbb{R}^{c_q \times c_{kv}}$, mask $S$ if needed
                        \STATE $\tilde{m} \gets \max(m^{(y, h)}, \text{rowmax}(S))$,  $P \gets \exp(S - \tilde{m})$, $\tilde{l} \gets \exp(m^{(y, h)} - \tilde{m})l^{(y, h)} + \text{rowsum}(P)$, $\tilde{C_o} \gets \exp(m^{(y, h)} - \tilde{m}) C_o^{(y, h)} + P C_v^{(x, h)}$
                        \STATE Store $C_o^{(y, h)} \gets \tilde{C_o}$, $l^{(y, h)} \gets \tilde{l}$, $m^{(y, h)} \gets \tilde{m}$
                    \ENDFOR
                \ENDFOR
            \ENDFOR
        \ENDFOR
        \STATE On chip, $B_o^{(r, i)} \gets \text{diag}(l)^{-1} B_o^{(r, i)}$
        \STATE Start DMA for sending $B_o^{(r, i)}$ to HBM
    \ENDFOR
\ENDFOR
\STATE Wait for all remaining $B_o$ and $U_{kv}$ DMA operations to complete
\RETURN Updated KV Cache and attention outputs
\end{algorithmic}
\end{algorithm}

\subsection{Tuning}

Tuning is an essential step for optimizing TPU kernels. RPA exposes four block-size parameters for tuning: $b_q$, $b_{kv}$, $c_q$, and $c_{kv}$. These parameters must be adjusted to account for variations across TPU generations and topologies, model characteristics (e.g., head dimension, context length), and workload distributions (e.g., sequence lengths and generation lengths as controlled by the scheduler). Specifically, $b_q$ and $b_{kv}$ are tuned primarily to improve overlap between DMA transfers and computation. In contrast, $c_q$ and $c_{kv}$ are tuned to reduce VREGs pressure and mitigate \textit{VREG spilling} by processing smaller compute blocks at a time.

RPA allows users to specify three distinct sets of block sizes corresponding to decode-only, prefill-only (with fixed-size chunk prefill), and mixed workloads. For decode-only workloads, tuning focuses on saturating memory bandwidth, as the computation is memory-bound. For prefill-only workloads, tuning aims to maximize MXU bundle utilization and minimize VREG spilling while maintaining high MFU, as this workload is compute-bound. Mixed workloads are more challenging to tune, but optimizations are possible if workload characteristics are partially known---for example, if the majority of $KV$ lengths are short while a small fraction are exceptionally long, tuning for the common short length can yield efficient execution.

We recommend performing tuning offline and storing the optimized block sizes in a lookup table, which can then be applied during pre-compilation at the server bring-up stage.

\subsection{Recompilation Handling}

Pallas kernels require JAX JIT compilation, which enforces static shapes and data types for all tensors. Any deviation from the expected shapes triggers a kernel recompilation. Importantly, such recompilation overhead is incurred on the critical serving path and therefore directly contributes to end-to-end latency. To prevent this, users of RPA must define two upper bounds during server bring-up: the \textbf{maximum number of tokens $s$} and the \textbf{maximum number of sequences $n$}. Input data must then be padded to these upper bounds along the corresponding dimensions to ensure that kernel shapes remain static and recompilation is never triggered.

\subsection{Quantization}

Currently, RPA supports per-tensor quantization, where each of $Q$, $K$, or $V$ shares a single scalar scaling factor. Supporting more granular quantization schemes, such as per-channel or per-head quantization, would require additional kernel modifications.

In the current design, if the number of $KV$ heads per device ($h_{kv}$) is smaller than the packing factor, RPA automatically inserts padding to satisfy tiling constraints during the preprocessing step. Users should consider this behavior when configuring data parallelism (DP) and tensor parallelism (TP) to ensure efficient memory utilization and avoid unnecessary padding overhead.

\section{Empirical Evaluation}

We evaluate RPA performance across three representative use cases: decode-only, prefill-only, and mixed workloads. For each scenario, we provide a detailed analysis of actual runtime performance using the Llama 3 8B model on TPU7x, and discuss best-use patterns as well as potential avenues for further optimization. Additionally, we include RPA performance measurements within vLLM TPU, illustrating the cumulative impact of our kernel-level optimizations.

\textbf{Benchmark Setting:} All experiments are conducted on TPU7x Ironwood (2 TensorCores). We evaluate multiple configurations, including varying attention head dimensions ($128$ and $256$) and the presence or absence of causal masking. Other model specifications are consistent with the Llama 3 8B model, comprising $32$ $Q$ heads, $8$ $KV$ heads, and \texttt{BF16} precision.

\begin{table}[ht]
\centering
\caption{TPU7x Ironwood (2 TensorCores per chip) Key Specifications}
\label{tab:tpu7x_specs}
\begin{tabular}{ll}
\hline
\textbf{Feature} & \textbf{Specification} \\ \hline
HBM Capacity per chip & 192 GB \\
Peak HBM Bandwidth per chip & 7380 GB/s \\
Peak BF16 Compute per chip & 2307 TFLOPs/s \\ \hline
\end{tabular}
\end{table}

\textbf{Ablation Study Setting:} To identify the dominant performance bottleneck (memory-bound vs. compute-bound) and evaluate whether $KV$ cache latency is effectively hidden by the RPA pipeline, we conduct a series of ablation experiments on the Llama 3 8B model on TPU7x by selectively disabling key components, including $KV$ cache updates, DMA operations, and FlashAttention (FA). Removing $KV$ cache updates isolates their contribution to overall latency, while disabling DMA operations approximates a compute-only regime, allowing us to characterize compute-bound behavior. Conversely, disabling FlashAttention isolates data movement, yielding a memory-bound regime dominated by DMA activity. This controlled decomposition enables us to attribute performance characteristics to individual components and assess the effectiveness of pipeline overlapping in hiding $KV$ cache update latency.

\subsection{Decode}
\subsubsection{Throughput and MBU}

In the decode-only setting, we schedule $n=128$ sequences per RPA invocation. All sequences are assigned an identical context length (i.e., $KV$ length), such that the maximum $KV$ length corresponds to the effective processing size for each sequence. This uniform configuration simplifies the analysis and enables consistent measurement of effective throughput under decode workloads.

We evaluate performance across varying context lengths and report both effective throughput and memory bandwidth utilization (MBU). The effective throughput is defined as:

$$ \text{Throughput (GB/s)} = \frac{n \cdot d_k \cdot [(\text{context length} + 1) \cdot 2h_{kv} + 2h_q] \cdot \text{data bytes}}{1024^3 \cdot \text{time}} $$

In this formulation, the context length accounts for $KV$ data fetched from both the $KV$ cache and newly projected $KV$. The additional term "+ 1" represents the $KV$ update written back from VMEM to the $KV$ cache in HBM. The factor $2h_{kv}$ corresponds to the merged $KV$ representation (i.e., combined $K$ and $V$ heads), while $2h_q$ accounts for both the input $Q$ and the output tensors, assuming the same data type for $Q, K, V,$ and the output.

MBU is calculated as the ratio of effective throughput to the theoretical peak bandwidth of the hardware:

$$ \text{MBU} = \frac{\text{Throughput}}{\text{Peak HBM Bandwidth}} \times 100\% $$

\begin{figure}[htbp]
    \centering
    \includegraphics[width=0.9\textwidth]{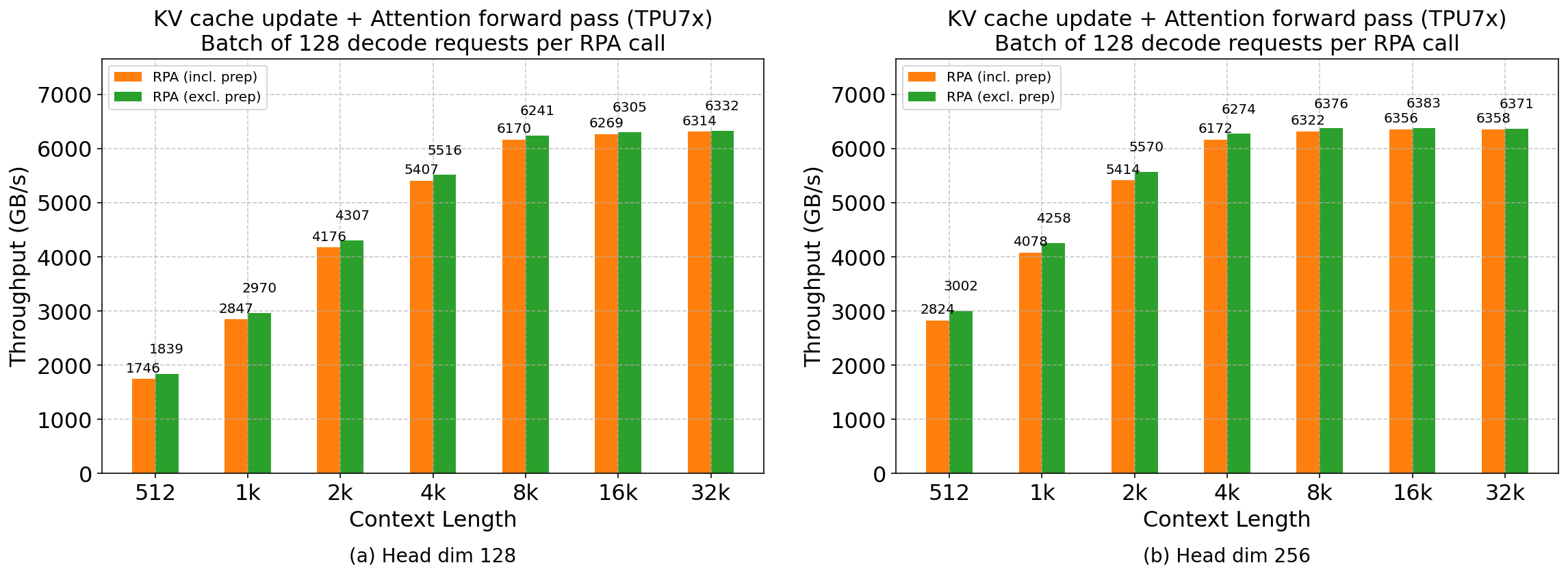}
    \caption{Effective Throughput of RPA (KV cache update + Attention forward pass) in decode.}
    \label{fig:decode_thpt}
\end{figure}

\begin{figure}[htbp]
    \centering
    \includegraphics[width=0.9\textwidth]{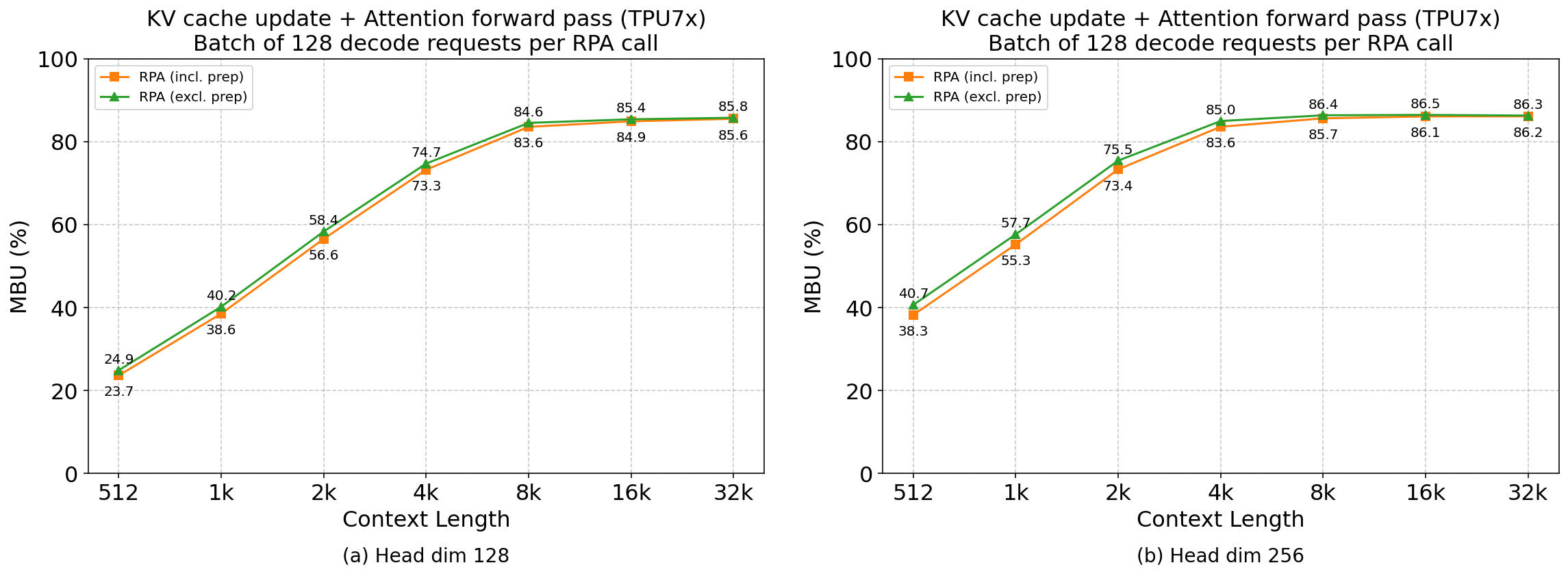}
    \caption{MBU of RPA (KV cache update + Attention forward pass) in decode.}
    \label{fig:decode_mbu}
\end{figure}

\subsubsection{Analysis of Decoding Results}

Several key observations can be drawn from the experimental results in Figure ~\ref{fig:decode_thpt} and Figure ~\ref{fig:decode_mbu}.

\textbf{Preprocessing Overhead.} The impact of preprocessing is relatively minor in our evaluation. This is largely due to the small batch size ($128$ tokens), where preprocessing overhead remains negligible compared to overall execution time. However, this overhead becomes more noticeable as the number of scheduled decode sequences increases (e.g., scaling to $1\text{K}$ sequences per RPA invocation). Additionally, when comparing configurations with head dimensions $128$ and $256$, we observe a modest increase in preprocessing cost in the latter case, as the tensor sizes---and consequently the data movement---are doubled.

\textbf{HBM Saturation and Payload Constraints.} For the standard Llama 3 8B configuration ($d_k=128$), we observe that at least an $8\text{K}$ context length is required to saturate HBM bandwidth. When the head dimension is increased to $256$, this requirement reduces to approximately $4\text{K}$. In both cases, this corresponds to fetching at least $16$~MB of $KV$ data per operation to fully utilize memory bandwidth.

However, such conditions are often not met in typical vLLM benchmarking setups, where decode workloads rarely reach these context lengths for all sequences. In mixed-batch scheduling, RPA relies on the maximum $KV$ length (as defined by the serving configuration), while the actual context lengths are often significantly shorter. 

One potential approach to address this inefficiency is to introduce \textit{mini-batching} across decode sequences. By starting asynchronous copies of $KV$ blocks for multiple requests, it becomes possible to increase the effective data movement per operation, thereby improving HBM bandwidth utilization before proceeding with computation.

\textbf{Summary of Efficiency.} Overall, when operating under conditions that sufficiently saturate memory bandwidth, RPA achieves up to $86\%$ MBU.

\subsubsection{Ablation Study}

\begin{figure}[htbp]
    \centering
    \includegraphics[width=0.9\textwidth]{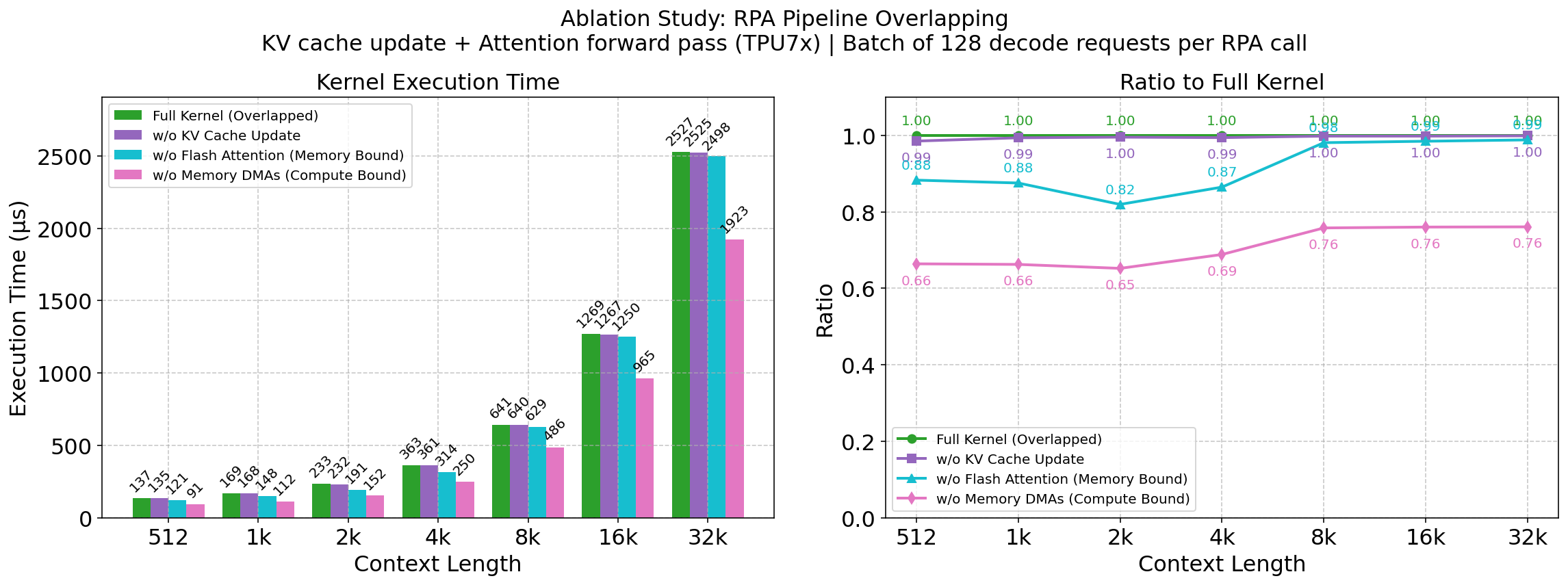}
    \caption{Ablation study of RPA pipeline in decode with Llama 3 8B.}
    \label{fig:decode_ablation}
\end{figure}

The ablation results indicate that KV cache update latency is negligible in the decode setting. This is because only a single token per sequence is inserted during each step, and the associated overhead is effectively hidden by computation through RPA’s pipeline overlapping. Furthermore, combined with prior observations, we find that once HBM bandwidth is saturated, RPA operates in a memory-bound regime, as removing FlashAttention does not lead to a noticeable change in overall latency. In contrast, prior to reaching HBM saturation, the overlap between computation and data movement is less effective, resulting in suboptimal resource utilization.

\subsection{Prefill}
\subsubsection{Speed and MFU}

In the prefill setting, we schedule a single sequence ($n=1$) per RPA invocation. The sequence length $s$ is varied from $512$ to $32\text{K}$ tokens. Under this configuration, the assigned sequence length directly corresponds to the effective processing length, and no padding is introduced. 

We evaluate performance across varying sequence lengths and report both compute Speed (in TFLOPs/s) and model FLOPs utilization (MFU). The definition of Speed depends on whether a causal mask is applied:

$$ \text{Speed}_{\text{causal=False}} \text{ (TFLOPs/s)} = \frac{4 \cdot s^2 \cdot h_q \cdot d_k}{10^{12} \cdot \text{time}} $$

$$ \text{Speed}_{\text{causal=True}} \text{ (TFLOPs/s)} = \frac{2 \cdot s \cdot (s + c_{kv}) \cdot h_q \cdot d_k}{10^{12} \cdot \text{time}} $$

When the causal mask is enabled, the FLOPs are not simply halved relative to the non-causal case. This is because each $KV$ compute block ($c_{kv}$) is still loaded and participates in matrix multiplication, even though only a subset of the results contributes to the final output due to masking. 

MFU is defined as the ratio between the achieved compute throughput (Speed) and the theoretical peak BF16 compute throughput of the hardware, adjusted for the maximum achievable MXU utilization. For example, a head dimension of $128$ may only utilize $50\%$ of the MXU, and this limitation is incorporated into the effective peak when computing MFU:

$$ \text{MFU} = \frac{\text{Speed}}{\text{Peak HW TFLOPs/s} \times \text{max MXU utilization}} \times 100\% $$

\begin{figure}[htbp]
    \centering
    \includegraphics[width=0.9\textwidth]{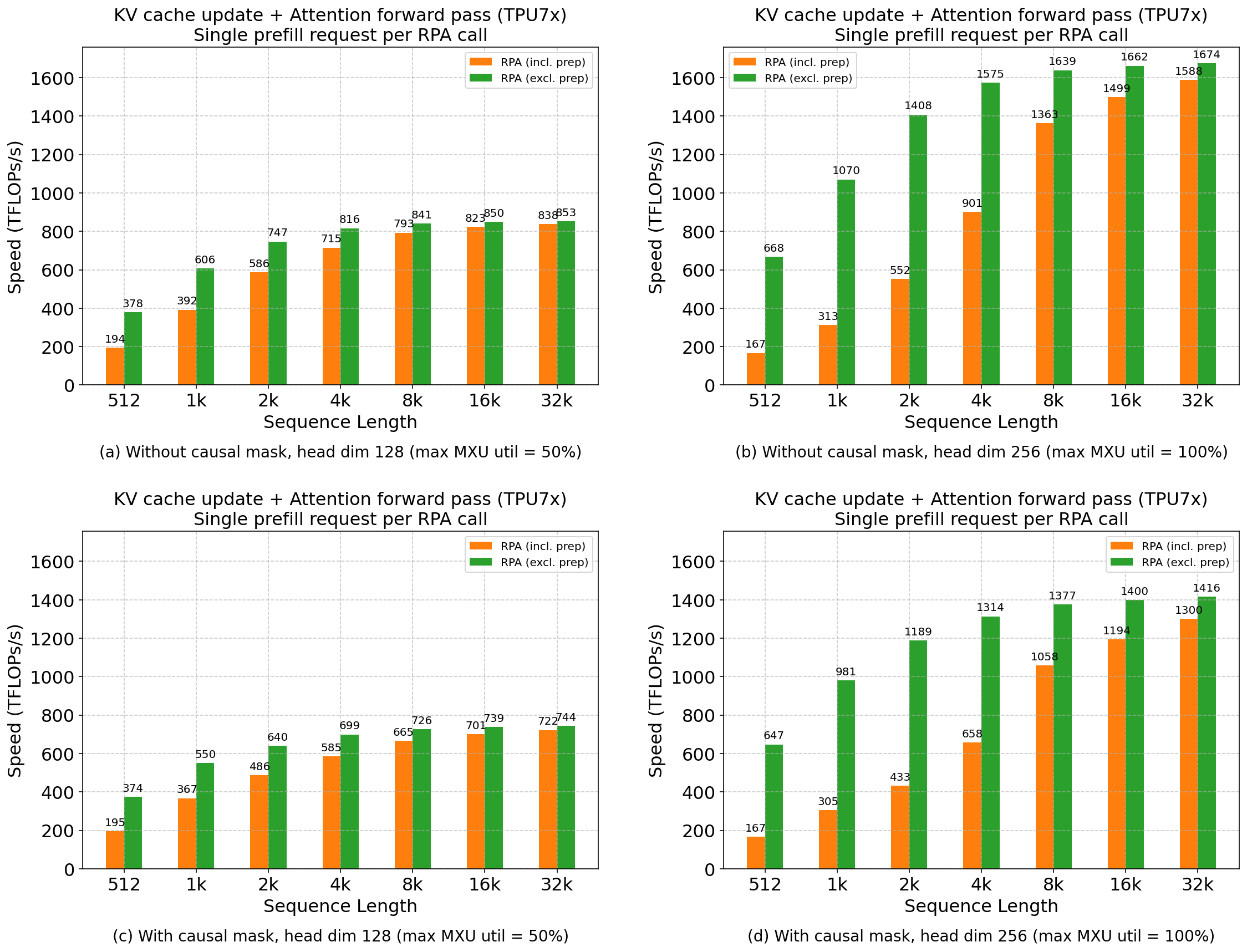}
    \caption{Speed of RPA (KV cache update + Attention forward pass) in prefill.}
    \label{fig:prefill_flops}
\end{figure}

\begin{figure}[htbp]
    \centering
    \includegraphics[width=0.9\textwidth]{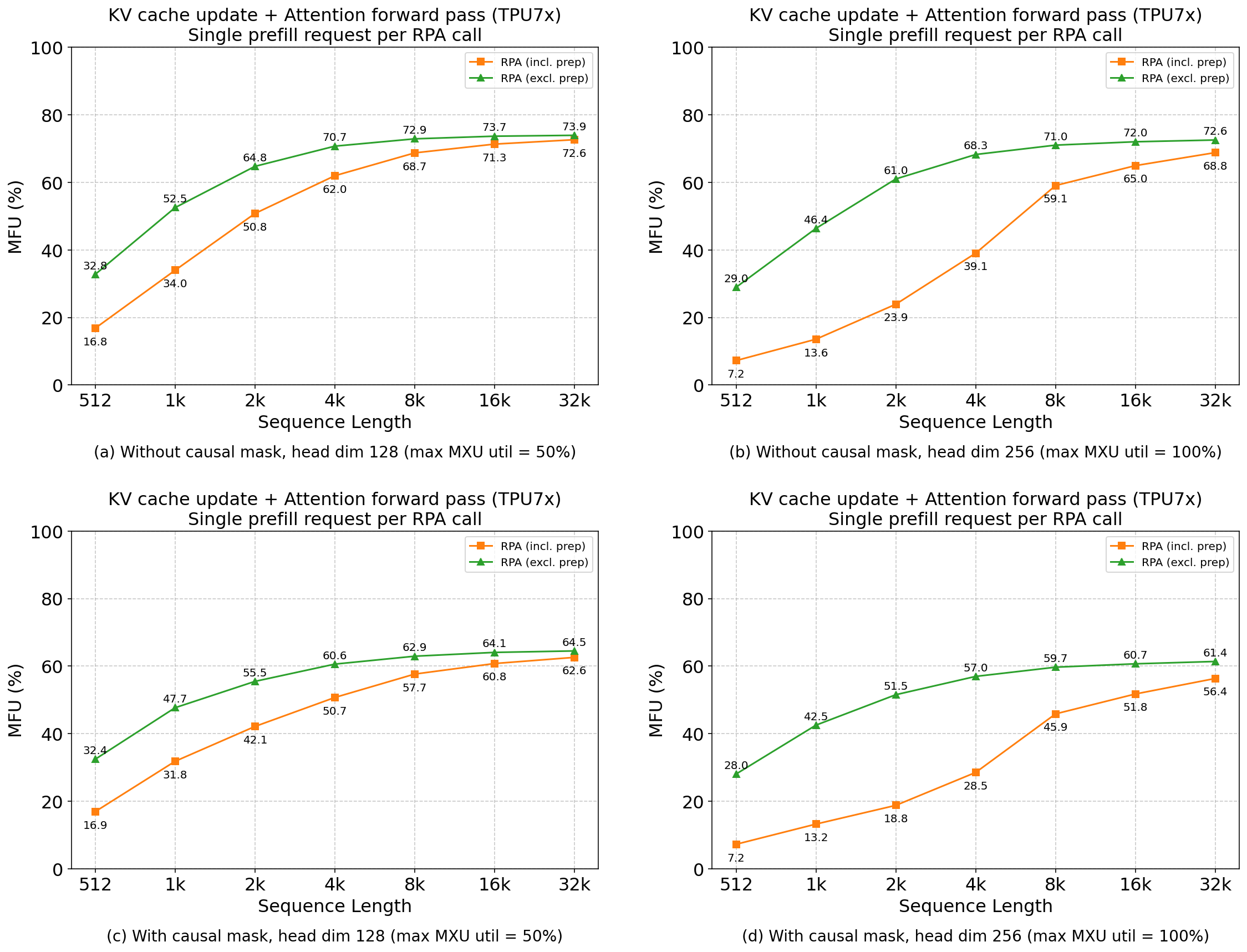}
    \caption{MFU of RPA (KV cache update + Attention forward pass) in decode.}
    \label{fig:prefill_mfu}
\end{figure}

\subsubsection{Analysis of Prefilling Results}
Several key observations can be drawn from the experimental results in Figure ~\ref{fig:prefill_flops} and Figure ~\ref{fig:prefill_mfu}.

\textbf{Preprocessing Overhead.}
In contrast to the decode setting, preprocessing has a more pronounced impact during prefill, as the sequence length $s$ directly determines the total number of tokens processed per invocation. This effect is more significant when the head dimension is increased from $128$ to $256$, due to the higher cost of transforming larger $Q$, $K$, and $V$ tensors. Nevertheless, once RPA reaches a compute-bound regime and saturates hardware compute throughput, the relative preprocessing overhead decreases, though it still accounts for approximately $2\%$--$8\%$ of total runtime.

A unique characteristic of the $d_k=256$ configuration (Figures 15b and 15d) is the inflection point observed at a sequence length of $8\text{K}$. For sequences below $8\text{K}$, the \textit{RPA (incl. prep)} curve exhibits a steep upward slope, indicating that the total execution time is dominated by preprocessing overhead. However, as the sequence length exceeds $8\text{K}$, the curve begins to decelerate and slightly decline. This transition marks the point where attention computation becomes the primary bottleneck. As discussed in Section 3.1.4, one potential approach to mitigate this overhead is to offload preprocessing offline; we leave this optimization to future work.

\textbf{Compute Saturation and Hardware Utilization}
For the standard Llama 3 8B configuration ($d_k=128$), we observe that a sequence length of at least $8\text{K}$ is required to saturate hardware compute throughput. A similar threshold applies to the $d_k=256$ configuration. However, increasing the head dimension to $256$ enables more efficient utilization of the TPU7x MXU, as attention computation can better match the native $256 \times 256$ MXU shape. As a result, the achievable peak FLOPs are effectively doubled compared to the $d_k=128$ case. 

This limitation is largely dictated by model architecture, as many existing open-source models adopt head dimensions smaller than $256$. Improving hardware utilization would therefore require co-design at the model level for TPU (e.g., adopting head dimensions that better align with MXU characteristics). 

In practical serving systems such as vLLM, prefill workloads often consist of multiple short sequences, each with lengths well below the $8\text{K}$ saturation threshold. As a result, individual sequences are unlikely to reach the compute-bound regime, leading to suboptimal hardware utilization. A natural approach to address this inefficiency is to introduce \textit{mini-batching} across short prefill sequences. By aggregating multiple sequences and staging their data from HBM to VMEM prior to execution, the system can increase the effective workload size and initiate computation across the batch simultaneously. This strategy raises operational intensity at earlier stages and enables higher utilization of hardware compute resources even when individual sequence lengths are small.

\textbf{Summary of Efficiency}
For the $d_k=128$ configuration, MFU is computed using an effective peak that accounts for the maximum achievable MXU utilization ($50\%$). Under compute-saturating conditions, RPA achieves up to $73\%$ MFU without causal masking and $63\%$ MFU with causal masking.

\subsubsection{Ablation Study}

\begin{figure}[htbp]
    \centering
    \includegraphics[width=0.9\textwidth]{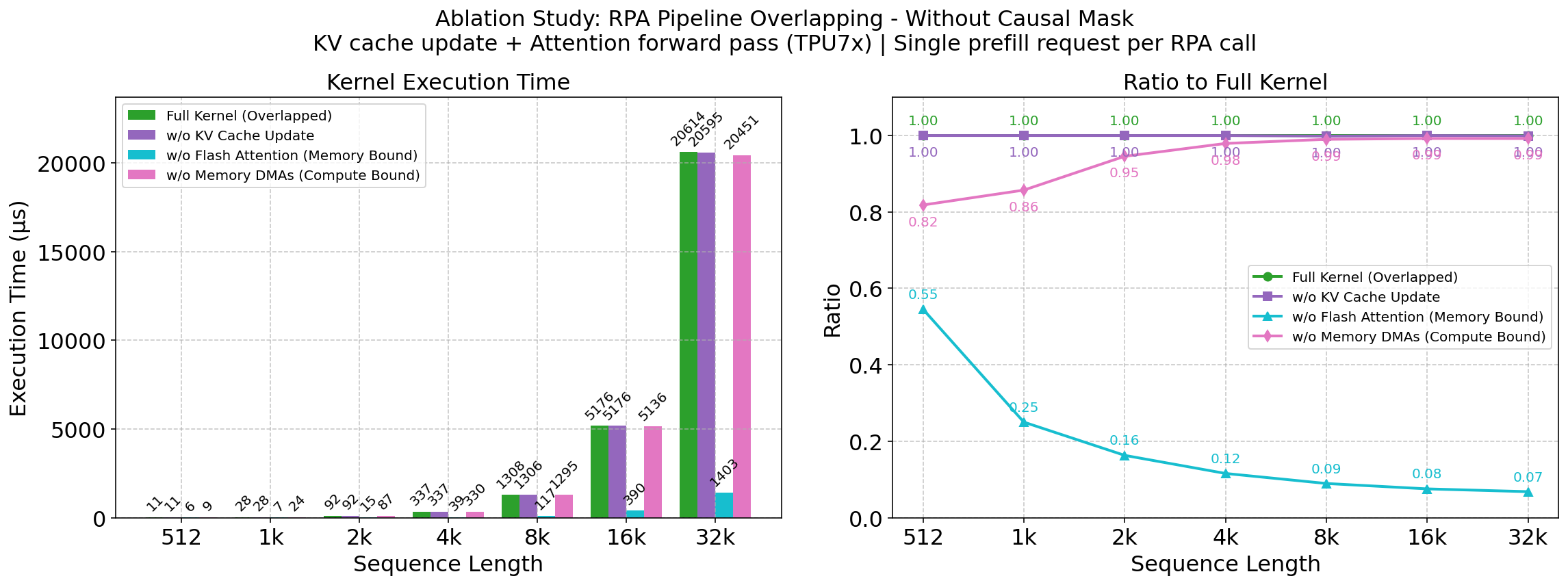}
    \caption{Ablation study of RPA pipeline in prefill (casual=False) with Llama 3 8B.}
    \label{fig:prefill_woc_ablation}
\end{figure}

\begin{figure}[htbp]
    \centering
    \includegraphics[width=0.9\textwidth]{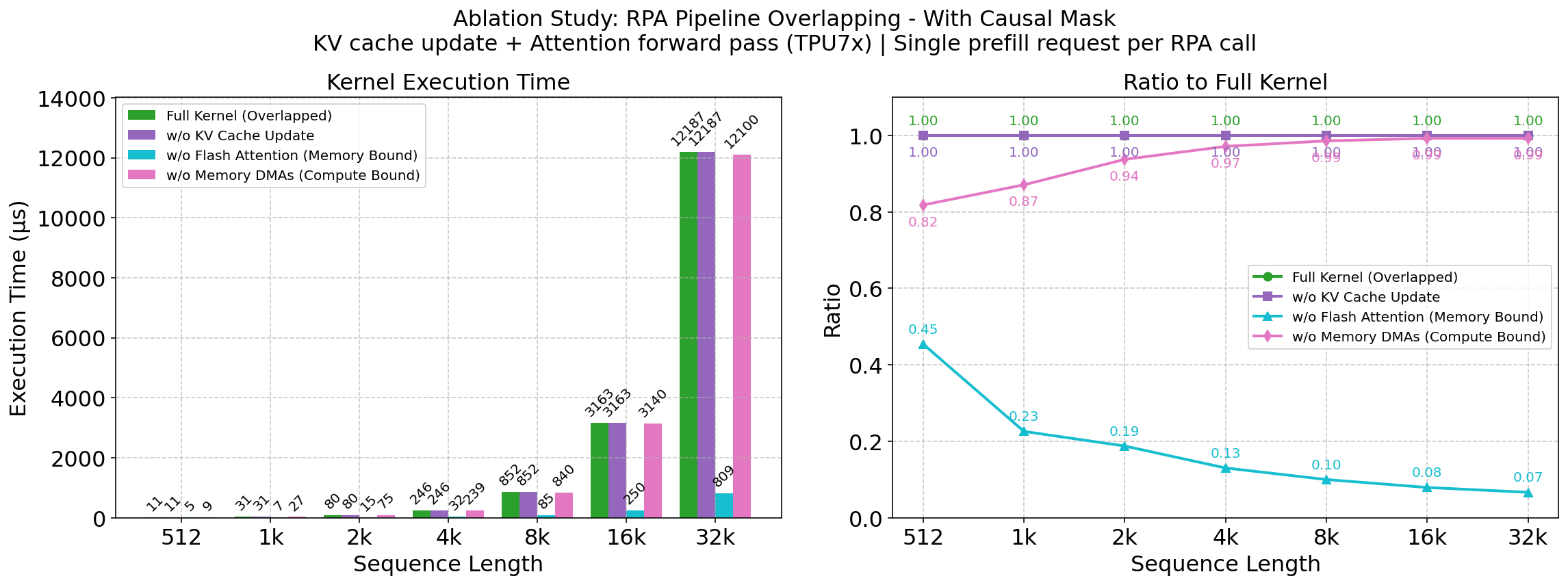}
    \caption{Ablation study of RPA pipeline in prefill (casual=True) with Llama 3 8B.}
    \label{fig:prefill_wc_ablation}
\end{figure}

The ablation results indicate that $KV$ cache update latency is largely hidden during the prefill phase. Unlike the decode setting, prefill exhibits substantially higher computational intensity, causing the workload to operate in a compute-bound regime across most configurations. As a result, the overhead associated with $KV$ cache updates is effectively overlapped with attention computation.

As the sequence length increases, the degree of compute-bound behavior becomes more pronounced. Once the sequence length reaches approximately $8\text{K}$, RPA is able to fully saturate the hardware BF16 compute throughput. At this point, all DMAs---including $KV$ cache fetch and update---are almost entirely masked by computation, as evidenced by the convergence between the compute-only curve and the full kernel execution time.

These results demonstrate that, in contrast to the decode phase where memory bandwidth can dominate, the prefill phase is predominantly compute-bound at scale. This enables effective overlap of communication and computation, leading to near-optimal hardware utilization.

\subsubsection{MXU Bundle Utilization}

\begin{figure}[htbp]
    \centering
    \includegraphics[width=0.9\textwidth]{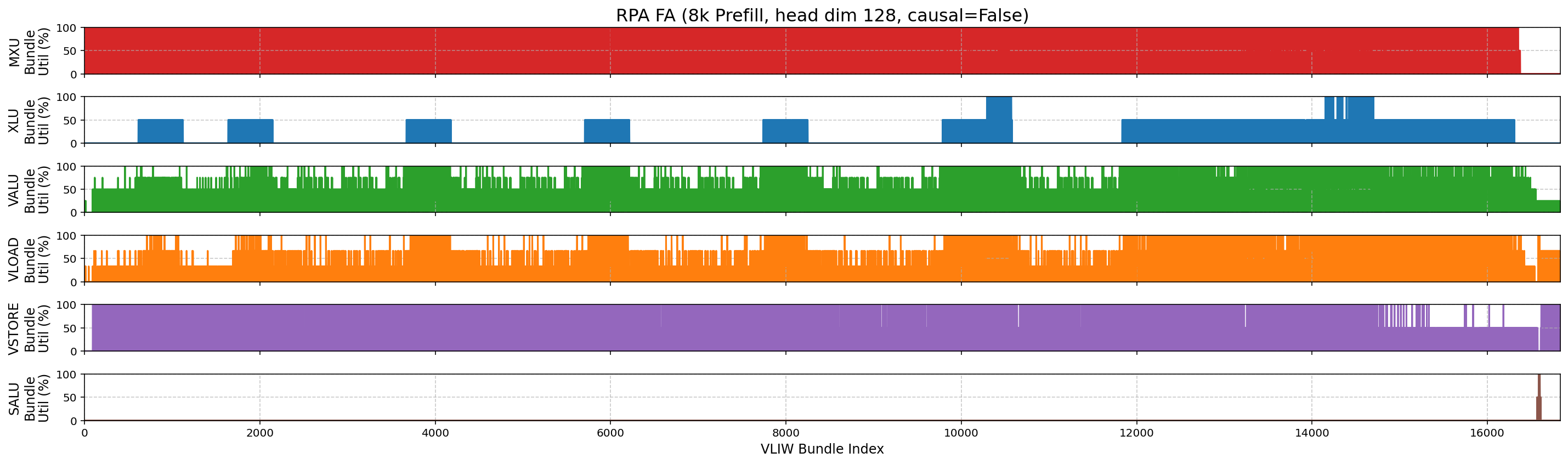}
    \caption{Bundle Utilization of FlashAttention in prefill (casual=False) with Llama 3 8B.}
    \label{fig:bundle_util_woc}
\end{figure}

\begin{figure}[htbp]
    \centering
    \includegraphics[width=0.9\textwidth]{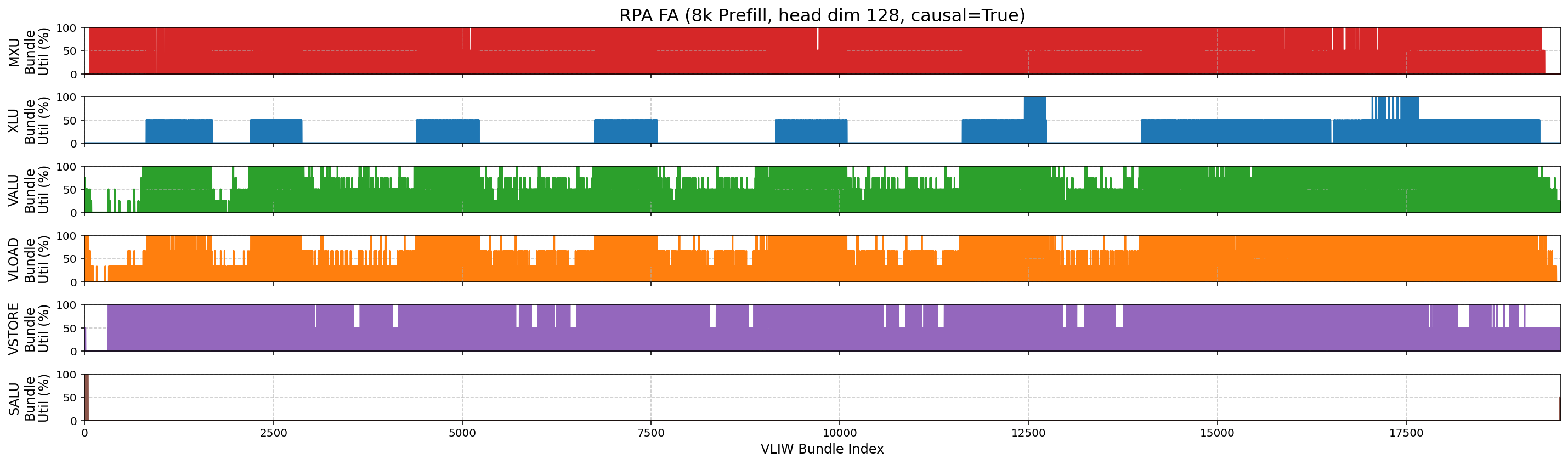}
    \caption{Bundle Utilization of FlashAttention in prefill (casual=True) with Llama 3 8B.}
    \label{fig:bundle_util_wc}
\end{figure}

TPU low-level operations (LLO) are scheduled using a VLIW (Very Long Instruction Word) architecture, where multiple independent operations are bundled and executed within a single cycle. An important indicator of compute efficiency is the utilization of MXU bundles---ideally approaching $100\%$---while auxiliary operations (e.g., VPU workloads) are effectively overlapped within the same execution window. 

Figure ~\ref{fig:bundle_util_woc} and ~\ref{fig:bundle_util_wc} profile the bundle utilization of the compute block, including FlashAttention, loading of $c_q$ and $c_{kv}$, separating $c_{kv}$ into $c_k$ and $c_v$, and storing the outputs. Across both configurations---without causal masking and with causal masking---RPA achieves near-peak MXU bundle utilization. This holds even for the Llama 3 8B configuration ($d_k=128$), where each MXU tile is effectively padded with zeros due to the mismatch with the native $256 \times 256$ systolic array. 

Despite the high MXU utilization, we observe that the Scalar ALU (SALU) slots remain largely underutilized during the compute phase. This suggests an opportunity for further optimization. In particular, scalar-intensive operations---such as page index computation for subsequent data-fetching stages---could be scheduled alongside the FlashAttention compute to better utilize SALU resources. We leave this optimization to future work.

\subsection{Mixed Batch}

\begin{figure}[htbp]
    \centering
    \includegraphics[width=0.9\textwidth]{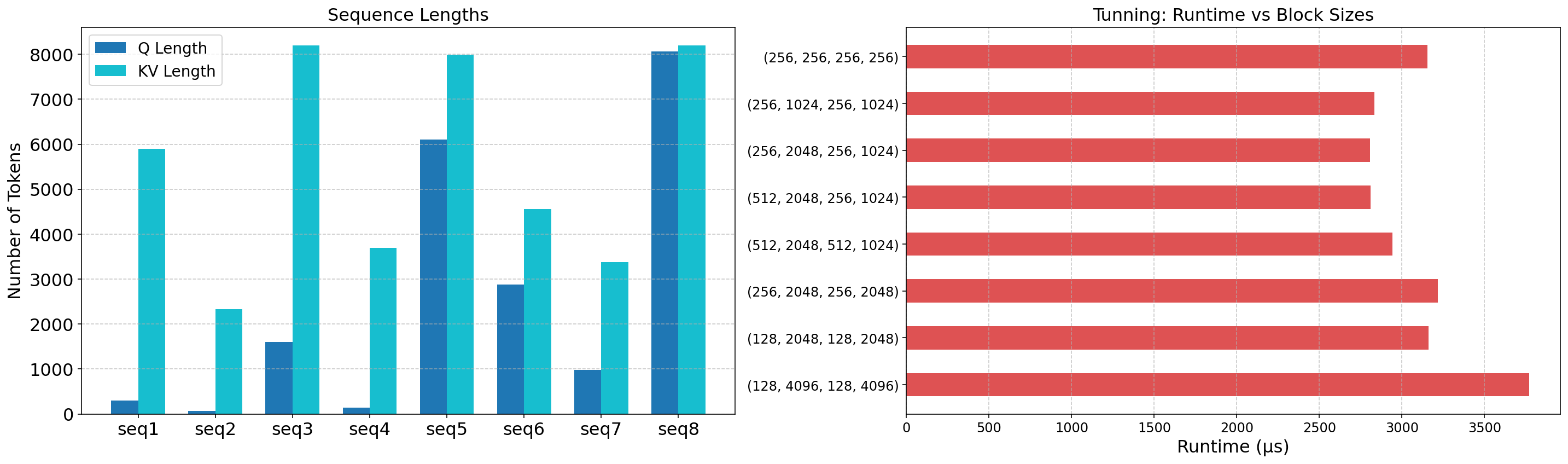}
    \caption{Example of a random mixed batch and block sizes tuning with Llama 3 8B.}
    \label{fig:mixed_tune}
\end{figure}

Mixed batches consist of sequences with ragged and dynamically varying lengths, where each sequence has a distinct context length ($KV$ length). RPA handles this variability through dynamic loops in the kernel, parameterized by block-size tuning parameters ($b_q, b_{kv}, c_q, c_{kv}$). These dynamic loops iterate over each sequence and issue DMA transfers with effective sizes to populate $B_q$ and $B_{kv}$ in VMEM, with any remaining capacity implicitly padded. Similarly, dynamic loops are used to load $c_q$ and $c_{kv}$ for computation. These dynamic loops enable the kernel to adapt to per-sequence variability at runtime.

However, this flexibility introduces inherent inefficiencies. Because computation must be aligned to fixed tile sizes $c_q$ and $c_{kv}$, the mismatch between unaligned sequence lengths and static compute sizes leads to unavoidable padding waste. This effect is particularly pronounced in highly ragged workloads, where the discrepancy between the actual token counts and the fixed tiling grid becomes significant.

As a result, there is no straightforward way to determine optimal block sizes solely from aggregate statistics, such as the maximum number of tokens $s$ and the maximum number of sequences $n$. Performance depends critically on the underlying distribution of sequence lengths within the batch, which is inherently non-deterministic in practical serving scenarios. Consequently, identical $(s, n)$ configurations can yield substantially different optimal tiling parameters depending on the specific workload distribution.

This sensitivity to input distribution remains a key challenge for RPA in mixed-batch settings (e.g., in vLLM), as tuning block sizes to achieve optimal performance under dynamic and heterogeneous sequence distributions is a non-trivial optimization problem.

\subsection{RPA in Production}

RPA currently serves as a critical component for the TPU backends of two prominent open-source inference frameworks: vLLM and SGLang. Since its integration into vLLM-TPU\cite{vllm_tpu_2025} in February 2025, RPA has enabled a $2\times$--$5\times$ increase in token throughput. As of October 2025, it has been officially adopted as the standard TPU backend for SGLang\cite{sglang_jax_2025}.

\begin{figure}[htbp]
    \centering
    \includegraphics[width=0.9\textwidth]{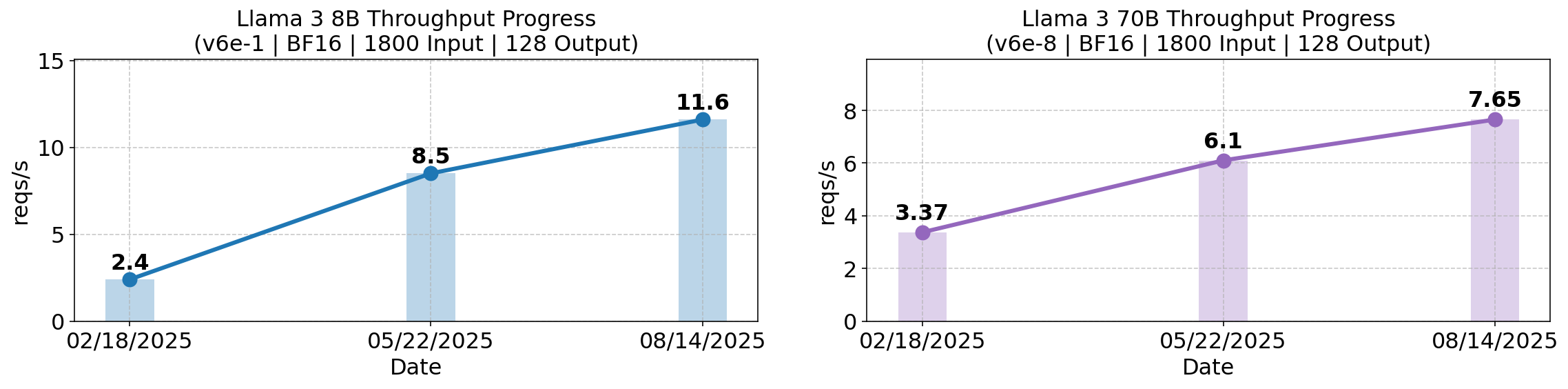}
    \caption{Llama 3 Throughput Evolution Powered by RPA in vLLM on TPU v6e.}
    \label{fig:vllm_progress}
\end{figure}

\section{Discussion and Future Directions}

RPA addresses the challenges of dynamism and raggedness that are difficult for traditional XLA-based approaches to efficiently compile and fuse---particularly for operations such as $KV$ cache updates (i.e., scatter) combined with attention. While RPA achieves high performance under hardware-saturated conditions and demonstrates flexibility across diverse workloads---including  decode-only,  prefill-only, and mixed-batch execution---it still presents several limitations and opportunities for further optimization.

\textbf{Preprocessing Overhead.} Preprocessing remains a non-negligible source of overhead, particularly in regimes where compute or memory bandwidth is not fully saturated. Potential directions to mitigate this cost include: (1) developing custom kernels for projection with tile constraints; (2) pursuing deeper kernel fusion to eliminate intermediate transformations; and (3) leveraging XLA layout propagation mechanisms, as discussed in Section ~\ref{sec:projection}, to reduce or eliminate explicit data reformatting.

\textbf{Mini-Batch Sequence Aggregation.} When hardware resources such as HBM bandwidth or compute throughput are underutilized, aggregating multiple sequences into a mini-batch within a single DMA or compute kernel can improve overall efficiency. By increasing the effective workload size per operation, this approach raises operational intensity and enables better utilization of both memory bandwidth and compute resources.

\textbf{Metadata Calculation Overlap}. Profiling indicates that SREGs are underutilized during the compute-intensive phases of FlashAttention, while metadata calculation for dynamic DMA addresses and sizes requires substantial scalar computation but leaves VREGs idle. To address this imbalance, we propose to precompute the metadata and stage it in SMEM within the attention loop, enabling overlap between scalar and vector execution and thereby hiding the associated latency.

\textbf{Scheduling Strategies.} Upstream scheduling plays a critical role in improving kernel efficiency. For example, generating fixed-size chunks for prefill workloads would reduce runtime dynamism, allowing the kernel to operate over more regular shapes. This, in turn, enables more efficient static allocation of VREGs and potential loop unrolling on TPU, improving overall execution efficiency.

\textbf{Disaggregated Serving.} RPA is well-suited for disaggregated serving architectures, where prefill and decode workloads are separated. Such separation enables more targeted optimization, as each workload exhibits distinct bottlenecks: decode is typically memory-bandwidth-bound, favoring optimizations that improve MBU, whereas prefill is often compute-bound, favoring higher MFU. Exploring RPA within disaggregated serving systems presents a promising direction for achieving more predictable and tunable performance.

\textbf{SparseCore Offloading.} Currently, the RPA architecture does not leverage SparseCore, as the custom kernel infrastructure for SC remains under active development. Future iterations of the kernel design aim to integrate SparseCore capabilities to further enhance performance and efficiency.

\section*{Acknowledgments}
We would like to express our deepest gratitude to the JAX and XLA teams at Google for the development of the Pallas and Mosaic infrastructure. This work would not have been possible without the robust foundation and continuous support of the TPU custom kernel stack. We extend special thanks to Adam Paszke, Sharad Vikram, and Blake Hechtman, who initiated the Pallas and Mosaic projects, as well as to the many XLA experts for their dedicated work in supporting the infrastructure that made the RPA kernel possible.

We also sincerely thank the vLLM and SGLang communities for their collaboration and support in integrating RPA into modern serving stacks. We would like to specifically recognize Brittany Rockwell and the technical leads who founded the v-team to build TPU-vLLM. Brittany’s work in bridging the gap between Google engineering and the open-source community was vital in shipping production-grade products and ensuring the success of TPU-based inference for both Google and the broader community.

Finally, we are profoundly grateful to Robert Hundt, David Majnemer, and Eli Bendersky for their rigorous technical reviews, insightful guidance, and steadfast support throughout the process of bringing this work to publication. We also thank our colleagues and the early adopters of RPA whose feedback helped refine the kernel into a production-ready solution for high-performance LLM deployment.

\bibliographystyle{plain} 
\bibliography{references} 

\appendix

\section{Full Benchmarking Results}

\begin{table}[H]
\centering
\small
\setlength{\tabcolsep}{6pt}
\caption{RPA Performance for Decode on TPU7x ($page\_size=256$, $n=128$, $d_k=128$)}
\label{tab:rpa_decode_performance}
\begin{tabular}{lcccccc}
\toprule
\textbf{Context} & \multicolumn{2}{c}{\textbf{Latency [$\mu$s]}} & \multicolumn{2}{c}{\textbf{Throughput [GB/s]}} & \multicolumn{2}{c}{\textbf{MBU [\%]}} \\
\cmidrule(lr){2-3} \cmidrule(lr){4-5} \cmidrule(lr){6-7}
\textbf{Length} & Incl. Prep & Excl. Prep & Incl. Prep & Excl. Prep & Incl. Prep & Excl. Prep \\
\midrule
512   & 145  & 137  & 1746 & 1839 & 23.66 & 24.91 \\
1024  & 176  & 169  & 2847 & 2970 & 38.58 & 40.24 \\
2048  & 240  & 233  & 4176 & 4307 & 56.58 & 58.36 \\
4096  & 370  & 363  & 5407 & 5516 & 73.27 & 74.75 \\
8192  & 649  & 641  & 6170 & 6241 & 83.61 & 84.57 \\
16384 & 1277 & 1269 & 6269 & 6305 & 84.94 & 85.43 \\
32768 & 2534 & 2527 & 6314 & 6332 & 85.56 & 85.80 \\
\bottomrule
\end{tabular}
\end{table}

\begin{table}[H]
\centering
\small
\setlength{\tabcolsep}{6pt}
\caption{RPA Performance for Decode on TPU7x (2 TC, $n=128$, $head\_dim=\textbf{256}$)}
\label{tab:tpu7x_2tc_performance}
\begin{tabular}{lcccccc}
\toprule
\textbf{Context} & \multicolumn{2}{c}{\textbf{Latency [$\mu$s]}} & \multicolumn{2}{c}{\textbf{Throughput [GB/s]}} & \multicolumn{2}{c}{\textbf{MBU [\%]}} \\
\cmidrule(lr){2-3} \cmidrule(lr){4-5} \cmidrule(lr){6-7}
\textbf{Length} & Incl. Prep & Excl. Prep & Incl. Prep & Excl. Prep & Incl. Prep & Excl. Prep \\
\midrule
512   & 178  & 168  & 2824 & 3002 & 38.27 & 40.68 \\
1024  & 246  & 236  & 4078 & 4258 & 55.26 & 57.71 \\
2048  & 370  & 360  & 5414 & 5570 & 73.35 & 75.48 \\
4096  & 649  & 638  & 6172 & 6274 & 83.63 & 85.02 \\
8192  & 1266 & 1256 & 6322 & 6376 & 85.67 & 86.39 \\
16384 & 2518 & 2507 & 6356 & 6383 & 86.13 & 86.49 \\
32768 & 5034 & 5023 & 6358 & 6371 & 86.15 & 86.33 \\
\bottomrule
\end{tabular}
\end{table}

\begin{table}[H]
\centering
\small
\setlength{\tabcolsep}{6pt}
\caption{RPA Performance for Non-Causal Prefill on TPU7x ($page\_size=256$, $n=1$, $d_k=128$)}
\label{tab:tpu7x_noncausal_performance}
\begin{tabular}{lcccccc}
\toprule
\textbf{Sequence} & \multicolumn{2}{c}{\textbf{Latency [$\mu$s]}} & \multicolumn{2}{c}{\textbf{TFLOPs/s}} & \multicolumn{2}{c}{\textbf{MFU [\%]}} \\
\cmidrule(lr){2-3} \cmidrule(lr){4-5} \cmidrule(lr){6-7}
\textbf{Length} & Incl. Prep & Excl. Prep & Incl. Prep & Excl. Prep & Incl. Prep & Excl. Prep \\
\midrule
512   & 22    & 11    & 194 & 378 & 16.86 & 32.76 \\
1024  & 44    & 28    & 392 & 606 & 33.94 & 52.50 \\
2048  & 117   & 92    & 586 & 747 & 50.78 & 64.76 \\
4096  & 384   & 337   & 715 & 816 & 61.98 & 70.70 \\
8192  & 1386  & 1308  & 793 & 841 & 68.78 & 72.88 \\
16384 & 5346  & 5176  & 823 & 850 & 71.32 & 73.66 \\
32768 & 20982 & 20614 & 838 & 853 & 72.68 & 73.98 \\
\bottomrule
\end{tabular}
\end{table}

\begin{table}[H]
\centering
\small
\setlength{\tabcolsep}{6pt}
\caption{RPA Performance for Non-Causal Prefill on TPU7x ($page\_size=256$, $n=1$, $d_k=\textbf{256}$)}
\label{tab:tpu7x_noncausal_hd256_performance}
\begin{tabular}{lcccccc}
\toprule
\textbf{Sequence} & \multicolumn{2}{c}{\textbf{Latency [$\mu$s]}} & \multicolumn{2}{c}{\textbf{TFLOPs/s}} & \multicolumn{2}{c}{\textbf{MFU [\%]}} \\
\cmidrule(lr){2-3} \cmidrule(lr){4-5} \cmidrule(lr){6-7}
\textbf{Length} & Incl. Prep & Excl. Prep & Incl. Prep & Excl. Prep & Incl. Prep & Excl. Prep \\
\midrule
512   & 51    & 13    & 167  & 668  & 7.26  & 28.95 \\
1024  & 110   & 32    & 313  & 1070 & 13.56 & 46.37 \\
2048  & 249   & 98    & 552  & 1408 & 23.91 & 61.03 \\
4096  & 610   & 349   & 901  & 1575 & 39.04 & 68.25 \\
8192  & 1613  & 1342  & 1363 & 1639 & 59.08 & 71.03 \\
16384 & 5867  & 5292  & 1499 & 1662 & 64.99 & 72.04 \\
32768 & 22160 & 21017 & 1588 & 1674 & 68.82 & 72.57 \\
\bottomrule
\end{tabular}
\end{table}

\begin{table}[H]
\centering
\small
\setlength{\tabcolsep}{6pt}
\caption{RPA Performance for Causal Prefill on TPU7x ($page\_size=256$, $n=1$, $d_k=128$)}
\label{tab:tpu7x_prefill_performance}
\begin{tabular}{lcccccc}
\toprule
\textbf{Sequence} & \multicolumn{2}{c}{\textbf{Latency [$\mu$s]}} & \multicolumn{2}{c}{\textbf{TFLOPs/s}} & \multicolumn{2}{c}{\textbf{MFU [\%]}} \\
\cmidrule(lr){2-3} \cmidrule(lr){4-5} \cmidrule(lr){6-7}
\textbf{Length} & Incl. Prep & Excl. Prep & Incl. Prep & Excl. Prep & Incl. Prep & Excl. Prep \\
\midrule
512   & 22    & 11    & 195 & 374 & 16.90 & 32.42 \\
1024  & 47    & 31    & 367 & 550 & 31.84 & 47.68 \\
2048  & 106   & 80    & 486 & 640 & 42.10 & 55.50 \\
4096  & 294   & 246   & 585 & 699 & 50.74 & 60.60 \\
8192  & 930   & 852   & 665 & 726 & 57.64 & 62.94 \\
16384 & 3332  & 3163  & 701 & 739 & 60.78 & 64.04 \\
32768 & 12556 & 12187 & 722 & 744 & 62.64 & 64.52 \\
\bottomrule
\end{tabular}
\end{table}

\begin{table}[H]
\centering
\small
\setlength{\tabcolsep}{6pt}
\caption{RPA Performance for Causal Prefill on TPU7x ($page\_size=256$, $n=1$, $d_k=\textbf{256}$)}
\label{tab:tpu7x_causal_hd256_performance}
\begin{tabular}{lcccccc}
\toprule
\textbf{Sequence} & \multicolumn{2}{c}{\textbf{Latency [$\mu$s]}} & \multicolumn{2}{c}{\textbf{TFLOPs/s}} & \multicolumn{2}{c}{\textbf{MFU [\%]}} \\
\cmidrule(lr){2-3} \cmidrule(lr){4-5} \cmidrule(lr){6-7}
\textbf{Length} & Incl. Prep & Excl. Prep & Incl. Prep & Excl. Prep & Incl. Prep & Excl. Prep \\
\midrule
512   & 51    & 13    & 167  & 647  & 7.23  & 28.06 \\
1024  & 113   & 35    & 305  & 981  & 13.22 & 42.52 \\
2048  & 238   & 87    & 433  & 1189 & 18.77 & 51.53 \\
4096  & 522   & 261   & 658  & 1314 & 28.51 & 56.97 \\
8192  & 1170  & 898   & 1058 & 1377 & 45.84 & 59.71 \\
16384 & 3912  & 3337  & 1194 & 1400 & 51.77 & 60.70 \\
32768 & 13951 & 12808 & 1300 & 1416 & 56.37 & 61.40 \\
\bottomrule
\end{tabular}
\end{table}

\begin{table}[H]
\centering
\footnotesize
\setlength{\tabcolsep}{6pt}
\caption{RPA Ablation Study for Decode on TPU7x ($page\_size=256$, $n=128$, $d_k=128$)}
\label{tab:ablation_latency}
\begin{tabular}{lccccc}
\toprule
\textbf{Context} & \multicolumn{2}{c}{\textbf{Latency [$\mu$s]}} & \multicolumn{3}{c}{\textbf{Ablated Latency [$\mu$s]}} \\
\cmidrule(lr){2-3} \cmidrule(l){4-6}
\textbf{Length} & Incl. Prep & Excl. Prep & w/o $KV$ Update & w/o FA & w/o DMA \\
\midrule
512   & 145  & 137  & 135  & 121  & 91   \\
1024  & 176  & 169  & 168  & 148  & 112  \\
2048  & 240  & 233  & 232  & 191  & 152  \\
4096  & 370  & 363  & 361  & 314  & 250  \\
8192  & 649  & 641  & 640  & 629  & 486  \\
16384 & 1277 & 1269 & 1267 & 1250 & 965  \\
32768 & 2534 & 2527 & 2525 & 2498 & 1923 \\
\bottomrule
\end{tabular}
\end{table}

\begin{table}[H]
\centering
\footnotesize
\setlength{\tabcolsep}{6pt}
\caption{RPA Ablation Study for Non-Causal Prefill on TPU7x ($page\_size=256$, $n=1$, $d_k=128$)}
\label{tab:ablation_noncausal_latency}
\begin{tabular}{lccccc}
\toprule
\textbf{Sequence} & \multicolumn{2}{c}{\textbf{Latency [$\mu$s]}} & \multicolumn{3}{c}{\textbf{Ablated Latency [$\mu$s]}} \\
\cmidrule(lr){2-3} \cmidrule(l){4-6}
\textbf{Length} & Incl. Prep & Excl. Prep & w/o $KV$ Update & w/o FA & w/o DMA \\
\midrule
512   & 22    & 11    & 11    & 6    & 9     \\
1024  & 44    & 28    & 28    & 7    & 24    \\
2048  & 117   & 92    & 92    & 15   & 87    \\
4096  & 384   & 337   & 337   & 39   & 330   \\
8192  & 1386  & 1308  & 1306  & 117  & 1295  \\
16384 & 5346  & 5176  & 5176  & 390  & 5136  \\
32768 & 20982 & 20614 & 20595 & 1403 & 20451 \\
\bottomrule
\end{tabular}
\end{table}

\begin{table}[H]
\centering
\footnotesize
\setlength{\tabcolsep}{6pt}
\caption{RPA Ablation Study for Causal Prefill on TPU7x ($page\_size=256$, $n=1$, $d_k=128$)}
\label{tab:ablation_prefill_latency}
\begin{tabular}{lccccc}
\toprule
\textbf{Sequence} & \multicolumn{2}{c}{\textbf{Latency [$\mu$s]}} & \multicolumn{3}{c}{\textbf{Ablated Latency [$\mu$s]}} \\
\cmidrule(lr){2-3} \cmidrule(l){4-6}
\textbf{Length} & Incl. Prep & Excl. Prep & w/o $KV$ Update & w/o FA & w/o DMA \\
\midrule
512   & 22    & 11    & 11    & 5   & 9     \\
1024  & 47    & 31    & 31    & 7   & 27    \\
2048  & 106   & 80    & 80    & 15  & 75    \\
4096  & 294   & 246   & 246   & 32  & 239   \\
8192  & 930   & 852   & 852   & 85  & 840   \\
16384 & 3332  & 3163  & 3163  & 250 & 3140  \\
32768 & 12556 & 12187 & 12187 & 809 & 12100 \\
\bottomrule
\end{tabular}
\end{table}


\end{document}